# Air-Sea Interactions on Titan: Lake Evaporation, Atmospheric Circulation, and Cloud Formation

Scot C. R. Rafkin[*] and Alejandro Soto

Department of Space Studies, Southwest Research Institute, 1050 Walnut Street, Suite 300, Boulder, CO, USA, rafkin.swri@gmail.com.

*Corresponding Author

Keywords:  Titan, Titan Seas, Titan Lakes, Titan Atmosphere, Air-Sea Interaction, Sea breeze, Titan Clouds

# Abstract

Titan's abundant lakes and seas exchange methane vapor and energy with the atmosphere via a process generally known as air-sea interaction.  This turbulent exchange process is investigated with an atmospheric mesoscale model coupled to a slab model representation of an underlying lake.  The impact of lake size, effective lake mixed layer depth, background wind speed, air-lake temperature differential, and atmospheric humidity on air-sea interaction processes is studied through dozens of two-dimensional simulations.  The general, quasi-steady solution is a non-linear superposition of a very weak background plume circulation driven by the buoyancy of evaporated methane with a stronger opposing thermally direct (sea breeze) circulation driven by the thermal contrast between the cold marine layer over the lake and the warmer inland air.  The specific solution depends on the value of selected atmosphere and lake property parameters, but the general solution of the superposition of these two circulations is persistent.  Consistent with previous analytical work of others, the sensible heat flux and the latent heat flux trend toward opposite and equal values such that their ratio, the Bowen ratio, approaches -1.0 in most, but not all, of the quasi-steady state solutions.  Importantly, in nearly all scenarios, the absolute magnitude of the fluxes trends toward very small values such that the equilibrium solution is also nearly a trivial solution where air-sea energy exchange is ~3 W m$^{-2}$ or less.   In all cases, a cool, moist, and statically stable shallow marine layer with nearly calm winds and small turbulent flux exchanges with a colder underlying lake is produced by the model.  The temperature of the lake, the marine properties of the air, and the strength of the sea breeze depends on the initial conditions and to a lesser degree, the boundary conditions.

# 1. Introduction

The global cycling of methane on Titan must be dominated on short climatological timescales by the exchange between the atmosphere, where the bulk of free methane is thought to reside, and the surface, which includes lakes, seas, and possibly a damp regolith [e.g., Atreya et al., 2006; Lunine and Atreya, 2008; Aharonson et al., 2009; Schneider et al., 2012].  While the amount of methane in the near-surface is uncertain [e.g., Zarnecki et al, 2005; Hayes et al., 2008; Mitchell et al., 2008; Atkinson et al., 2010; Turtle et al., 2011; Hamelin et al., 2012; Turtle et al., 2018] there is no doubt that Titan's surface, particularly the northern high latitudes at present, is covered with lakes that are likely to contain substantial amounts of liquid methane [Cordier et al., 2009; Cordier et al., 2012; Lorenz et al., 2014].  These lakes serve as the only contemporary, persistent, and known source of methane to the atmosphere.

The exchange of sensible heat and latent heat (i.e., evaporation) between the lakes and the atmosphere is accomplished through turbulent fluxes. The study of the turbulent exchange process falls under the topical umbrella of air-sea interaction, which has a deep and rich history in the terrestrial literature [e.g., Bjerkenes, 1964; Pond, 1971; Hsu, 2005; Bishop et al., 2017].   On Earth, air-sea interaction plays a significant, if not dominant, role in phenomena that include sea and lake breezes [Pielke et al., 1974; Smith, 1988; Crosman and Horel, 2010], cloud formation [Oliver et al., 1978; Kingsmill et al., 1995; Miller et al., 2003], lake effect snowstorms [Lavoie, 1972; Chou and Atlas, 1982], the El Nino/Southern Oscillation [Emanuel, 1987; Zebiak, 1993; Alexander et al., 2002], and tropical cyclone development [Emanuel, 1986; Wu et al., 2005; Black et al., 2007]. Although sea breezes and ENSO-like climate oscillations have yet to be detected or observed on Titan, the presence of cloud features have been hypothesized to form as a result of sea breeze and local lake evaporation [Brown et al., 2009a; Brown et al., 2009b]. General circulation modeling studies of Titan have shown lakes have an impact on polar meteorology [Tokano, 2009] and the overall global distribution of methane [Lora and Ádámkovics, 2017]. Rafkin and Barth [2015] determined that it was primarily the amount of boundary layer methane rather than temperature variations that produced an environment conducive to deep convective cloud formation. Thus, variations in the global atmospheric methane distribution are ultimately manifested as clouds far from the only known permanent methane source [Rannou et al., 2006; Rodriguez et al., 2009] and drive deep convection [Mitchell et al., 2009], which provides a mechanism by which evaporated methane is returned to the surface [Tokano, 2011; Hueso and Sanchez-Lavega, 2006; Rafkin and Barth, 2015]. Evaporation from a surface moistened by prior rain can also contribute to the humidification of the atmosphere, but the methane in that rain is likely to have originated from the lakes at some point. Even tropical cyclones, driven by air-sea interaction, cannot be ruled out on Titan [Tokano et al., 2013].

The importance of air-sea interaction on Titan may extend beyond meteorological features. Sea breezes locally modify the large-scale background wind. Over lakes, the wind perturbations can disturb the lake surface, which is manifested as waves, and winds can drive internal circulations and convection that chemically and thermally mix the lake [Tokano and Lorenz, 2015; Tokano and Lorenz, 2016]. Climatological mean wind speeds at the surface are generally thought to be very weak and unlikely to force waves on a regular basis [Lorenz et al., 2005; Lorenz et al., 2010; Lorenz and Hayes, 2012; Hayes et al., 2013]. Radar observations of Titan lakes generally revealed them to be smooth, if not glassy [Wye et al., 2009; Grima et al., 2017], and the Cassini Visible and Infrared Mapping Spectrometer (VIMS) observed specular reflections that imply little to no lake roughness [Stephan et al., 2010; Barnes et al., 2011; Zebker et al., 2014]. On the other hand, evidence for some wave activity was found [Barnes et al., 2014], and waves are one possible explanation for "magic islands" (ephemeral and localized radar-derived roughness) in Ligeia Mare [Hofgartner et al., 2016]. Sea breezes may be the key to understanding lake roughness observations and may be important for understanding chemical and energy cycling.

Future observations and missions could better constrain air-sea interactions on Titan. The Titan Mare Explorer (TiME) previously proposed under the NASA Discovery Program would have placed a floating spacecraft in Ligeia Mare [Stofan et al., 2013]. Refined expectations of the meteorology and the air-sea interactions that this spacecraft would have directly experienced and measured would likely have benefited that mission design [Lorenz et al., 2012; Lorenz and Mann, 2015; Lorenz, 2015], as it will for any future Titan lake explorers. While the New Frontiers Titan Dragonfly mission [Lorenz et al., 2018] recently selected by NASA will reconnoiter the tropics far from any seas, the physics of air-sea interaction provides useful guidance for interpreting meteorological conditions over damp ground should the dunes within Selk Crater turn out to be so.

Air-sea interaction has not been widely investigated on Titan. Tokano (2009) investigated the seasonal changes of lakes using a general circulation model (GCM), but did not provide any meaningful discussion of the lake-scale atmospheric circulations that result from air-sea exchange. The use of a GCM model completely excludes the resolution of lake/sea breezes, and while the fluxes in the GCM are parameterized in a manner similar to this study, the mesoscale circulations that result from the flux exchange cannot be captured in the GCM. Mitri et al. [2007], hereafter M07, is presently the most comprehensive attempt to model the air-sea interaction process. Using a 1-D analytical model of the lake-atmosphere system, M07 found relatively large lake evaporation rates—sufficient to reduce lake levels by ~0.3 to 10 m over a season in the absence of any precipitation or subsurface resupply. As a consequence of the large evaporation rates (~0.3 × 10$^3$ to 5 × 10$^3$ kg m$^{-2}$ yr$^{-1}$), M07 also determined that evaporation from lakes alone was sufficient, by orders of magnitude, to resupply atmospheric methane against photochemical destruction.

The fundamental principle behind the M07 model is that subsaturated air blowing over a lake will induce evaporation until a balance condition of sensible and latent heat flux is established. Evaporation will cool the lake and moisten the atmosphere. As the lake (but not the atmosphere) cools, the magnitude of the sensible heat flux increases in proportion to the difference in temperature between the air and lake. At the same time, the moistening atmosphere and cooling lake (and therefore the saturation vapor pressure over the lake) drives down the magnitude of latent heat flux. The sign of the sensible heat flux is defined here as positive when energy is transferred from the sea to the air (a gain in energy by the air), and the latent heat flux is defined as negative when there is evaporative cooling of the lake (a loss of energy from the air to the sea). When the two fluxes are equal and opposite, air-sea interaction is assumed to reach a flux equilibrium state through which the lake temperature and fluxes may be diagnosed given a constant initial air temperature. Importantly, the equilibrium flux condition does not necessarily imply a thermodynamic or vapor equilibrium (i.e., the lake and atmosphere are not the same temperature nor is the air saturation). Note that the latent heat flux in SI units (Wm$^{-2}$) can be converted to an evaporation rate expressed as m s$^{-1}$ by dividing the flux by $L_v \rho_{l,CH_4}$ with the values given in Table 1.

The goal of this paper is to advance the understanding of air-sea interaction on Titan through the use of an atmospheric model that explicitly captures many of the complex, non-linear feedbacks between the atmosphere and underlying liquid reservoir. We seek to establish the nature and dynamics of atmospheric circulations driven by air-sea interaction, the thermodynamic structure of the atmosphere and its potential for cloud formation, and the potential for generating waves that would have been observable by Cassini or future orbiters. We further seek to evaluate and test assumptions of prior models of Titan air-sea interaction, to provide an update on global lake evaporation rates, and to introduce a useful and developing model tool for characterizing the marine environment in support of future Titan concepts and missions.

## 2. Model Description

We have developed a Titan mesoscale model based on the National Center for Atmospheric Research's (NCAR) Weather Research and Forecasting model (WRF), which we call mtWRF (mesoscale Titan WRF)[1]. WRF is an atmospheric model that is regularly used for terrestrial meteorological research and numerical weather prediction. We use the Advanced Research WRF (ARW) dynamical core as the basis for mtWRF. This dynamical core solves the fully compressible, Eulerian, non-hydrostatic fluid equations on an Arakawa C horizontal grid and a modified sigma pressure vertical coordinate [Skamarock, et al., 2008]. For our investigation, we used mtWRF in a two-dimensional configuration, i.e., one horizontal dimension and the vertical dimension. This configuration is consistent with prior analytical studies (e.g., M07) and is sufficient to study the essence of air-sea interaction physics.

---

[1] mtWRF is not related to the TitanWRF model, except that both models are derived from the same terrestrial NCAR WRF model. Otherwise, the development of mtWRF and TitanWRF are independent.

The radiative time constant in Titan's lower troposphere is much longer than a Titan year, therefore there is only a very weak diurnal signal in the atmosphere [Tomasko et al. 2008]. Although there is a diurnal signal in the ground temperatures, this signal is primarily in the tropics and over the Titan year. The polar regions, where the lakes are located, have almost no diurnal signal in the ground temperature [Tokano 2005]. For these reasons, the mtWRF simulations shown here do not use a radiative transfer scheme, although implementation of such a scheme is planned for the future.

We set the atmospheric and surface constants to match those of Titan. Specifically, the atmospheric composition and related thermodynamic properties were changed to match those of a nitrogen atmosphere with methane. The gravity and planetary radius were also changed to Titan values. Table 1 shows the various model parameters used in the mtWRF model.

Table 1. Model Parameters

| Symbol | Unit | Value | Parameter | Ref. |
|---|---|---|---|---|
| $m_{N_2}$ | amu | 28.0134 | Atomic weight of $N_2$ | 2 |
| $m_{CH_4}$ | amu | 16.042646 | Atomic weight of $CH_4$ | 2 |
| $m_{dry}$ | g/molecule | 28.67 | Mean molecular weight of air | 1 |
| $R$ | J mol$^{-1}$ K$^{-1}$ | 8.3145 | Molar gas constant for Titan | 2 |
| $R_d$ | J kg$^{-1}$ K$^{-1}$ | 290.0 | Dry air gas constant for Titan | 1 |
| $R_v$ | J kg$^{-1}$ K$^{-1}$ | 518.275 | Methane vapor air gas constant ($R/m_{CH_4}$) | 2 |
| $c_{p,air}$ | J kg$^{-1}$ K$^{-1}$ | 1044.0 | Specific heat of Titan's air | 1 |
| $c_{p,v}$ | J kg$^{-1}$ K$^{-1}$ | 1950 | Methane vapor specific heat, constant pressure | 3 |
| $c_{v,v}$ | J kg$^{-1}$ K$^{-1}$ | 1431.71 | Methane vapor specific heat, constant volume ($c_{p,v} - R_v$) | 2,3 |
| $c_{p,l}$ | J kg$^{-1}$ K$^{-1}$ | 3379 | Methane liquid specific heat, constant pressure @ 94 K | 5 |
| $L_v$ | J kg$^{-1}$ | 5.1x10$^5$ | Enthalpy of vaporization of methane at 112 K | 6,7 |
| $\rho_a$ | kg m$^{-3}$ | 5.5759 | Density of the atmosphere at 94 K | 5,10 |
| $\rho_{l,CH_4}$ | kg m$^{-3}$ | 447 | Density of liquid methane at 94 K | 4 |
| $P_0$ | Pa | 1x10$^5$ | Reference pressure | |
| $T_0$ | K | 100 | Reference temperature | |
| $g$ | m s$^{-2}$ | 1.352 | Gravity at Titan's surface | 2 |
| $\nu_a$ | m$^2$s$^{-1}$ | 1.18x10$^{-6}$ | Viscosity (momentum diffusivity) of Titan's atmosphere | 9 |
| $\alpha_a$ | m$^2$ s$^{-1}$ | 1.5 x10$^{-6}$ | Thermal diffusivity of the atmosphere | 5 |
| $D_{CH_4}$ | m$^2$ s$^{-1}$ | 2.05 x 10$^{-6}$ | Methane vapor diffusivity | 5 |
| $k_{l,CH_4}$ | W m$^{-1}$ K$^{-1}$ | 0.2 | Thermal conductivity of methane liquid | 5 |
| $k_a$ | W m$^{-1}$ K$^{-1}$ | 0.01 | Thermal conductivity of the atmosphere | 5,8,10 |

The atmospheric surface layer is the layer where there are turbulent exchanges between the surface and atmosphere.  mtWRF uses Monin-Obukhov (MO) similarity theory as described by Janjić [1990, 1994] to calculate surface fluxes; this is the Mellor-Yamada-Janić (MYJ) physics option in WRF [Skamarock et al., 2008]. The MO paramaterization solves for the bulk aerodynamic coefficient, $C_D$, in the bulk aerodynamic flux formulation:

$$\rho_a \overline{w'\chi'} = C_D \|\vec{V}\|(\chi_a - \chi_{sfc}) \qquad (1)$$

where $\rho_a$ is atmospheric density, $\overline{w'\chi'}$ is the Reynolds averaged correlation of subgrid vertical velocity ($w$) and the quantity of interest ($\chi$), $\vec{V}$ is the horizontal wind vector at the roughness height, and $\chi_a - \chi_{sfc}$ is the difference between the value of the quantity in the atmosphere and surface. $C_D$ is a function of subgrid scale turbulence, as diagnosed through the Bulk Richardson Number, which itself is a function of atmospheric stability and wind shear. Fluxes will tend to be large if turbulence is large (e.g., an unstable atmosphere or large wind shear), if wind speed is large, or if there is a large difference between the atmosphere and surface property. In contrast, a stable atmosphere, low wind speeds, or similar atmospheric and surface properties will result in a reduced flux.

Because of the lower Titan gravity, the relationship between the roughness length, $z_0$, and the friction velocity, $u_*$, are modified from the base WRF code. In Janjic [1994], the roughness length is calculated as:

$$z_0 = \frac{0.11 \nu_a}{u_*} + \frac{0.018 u_*^2}{g} \qquad (2)$$

where $\nu_a$ is the viscosity (momentum diffusivity) of the atmosphere and $g$ is the gravity. Fig. 1 shows the difference between the friction velocity generated roughness length on Titan versus Earth. On Titan, the roughness length is larger for nearly all values of friction velocity.

Traditionally, the atmospheric surface layer over a liquid body has been partitioned into three regimes [Janjic, 1994]: (1) a smooth and transitional regime, (2) a rough regime, and (3) a rough regime with spray.

The transitions between these regimes, determined by a roughness Reynolds number ($Rr = z_0 u_*/\nu$), represent an increase in the roughness and wave activity of lake and ocean surfaces. In the Earth WRF model, the smooth/transitional regime occurs for u* < 0.225m/s. The rough regime occurs for 0.225 m/s < u* < 0.7 m/s. And, finally, the rough regime with spray occurs for u* > 0.7 m/s. These values are similar to the ones suggested by Janjic [1994]. As Cassini observations have only seen smooth to possibly slightly perturbed lake surfaces (Grima et al. 2017), we have applied only the smooth and transitional portion of the atmospheric surface layer calculations. This configuration is appropriate for the available data and the type of initial mesoscale investigation that we are conducting.

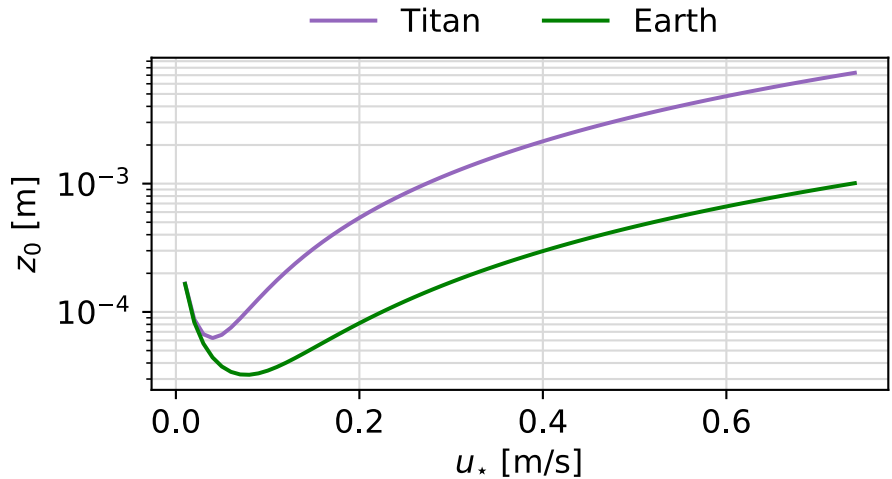

Figure 1. The relationship between $z_0$ and $u_*$ for Titan and Earth, based on the formulation from Janjic [1994]. The line for Earth matches Figure 3 in Janjic [1994].

Calculating the exchange of fluxes between the atmosphere and the surface for both lakes and land requires values for the molecular diffusivities of momentum, heat, and methane vapor. The MYJ scheme uses diffusivities that are constant with respect to temperature and pressure. The NIST Chemistry Webbook provides the atmospheric viscosity of a pure nitrogen atmosphere for a range of temperatures and pressures [Lemmon et al. 2017]. Fig. 2 shows the momentum diffusivity (i.e., kinematic viscosity), the thermal diffusivity, and the methane vapor diffusivity as a function of temperature from 90 K to 100 K. Similarly, the thermal diffusivity was derived from the thermal conductivity provided by the NIST Chemistry Webbook. Finally, the vapor diffusivity for methane in Titan's atmosphere was derived using an equation from Graves et al. [2008] and Lorenz [1993]. The surface layer physics in mtWRF use a constant value for each of the diffusivities, and the values at 94 K are chosen for mtWRF, as shown in Table 1 and Fig. 2. Fixing the value is done for simplicity, as the variations over the temperature ranges considered here are small. Future versions of mtWRF may introduce a temperature dependent value.

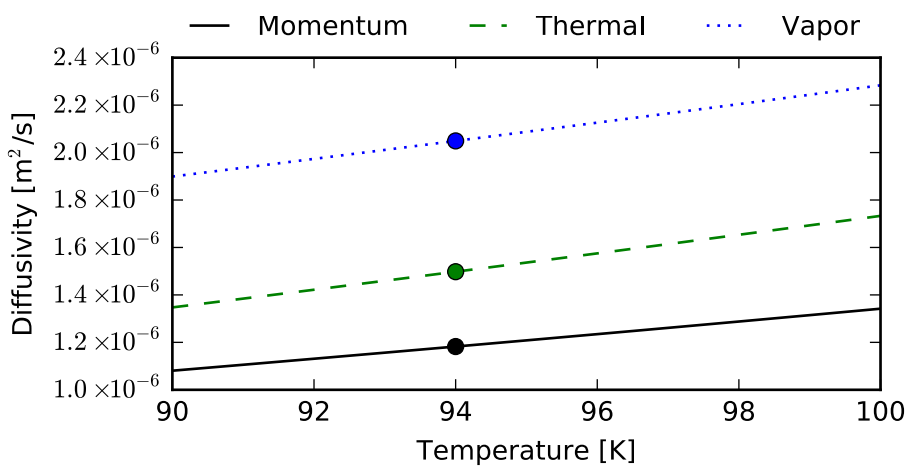

*Figure 2. The momentum, thermal, and vapor diffusivities of Titan's atmosphere as a function of temperature. The circle represent the values used in mtWRF. The value of the fixed diffusivities are within 10% the range of values expected for typical Titan surface temperatures.*

The surface fluxes of heat and moisture generated by the atmospheric surface layer scheme are the lower boundary conditions for the planetary boundary layer (PBL) scheme that parameterizes subgrid scale atmospheric eddy diffusion. The model uses the Mellor-Yamada-Janjic (MYJ) implementation to calculate turbulent kinetic energy vertical mixing in the PBL [Skamarock et al., 2008].

We included a slab lake model in mtWRF to account for a cooling lake on Titan. This slab model assumes that thermodynamic and energetic balance occurs instantaneously over a single layer in the lake with a thickness defined as the mixed layer depth, D. Since radiative transfer is assumed unimportant over the mesoscale simulation timescale, the slab lake model depends only on mixed layer depth, sensible heat flux, and latent heat flux, which form a prognostic lake temperature equation:

$$\frac{dT_L}{dt} = -\frac{1}{c_L \rho_L D} \left( c_A \rho_A \overline{w'T'} + L \rho_A \overline{w'q'} \right) \quad (2)$$

where $T_L$ is the lake temperature, t is time, $c_L$ is the lake specific heat, $\rho_L$ is the lake density, D is the mixed layer depth of the lake, $c_A$ is the atmospheric specific heat, $\rho_A$ is the atmospheric density, w is the vertical velocity, T is the atmospheric temperature, L is the enthalpy of vaporization, and q is the specific humidity. Fig. 3 shows the lake cooling rates achieved for a range of lake mixed layer depths and heat fluxes $\left( c_A \rho_A \overline{w'T'} + L \rho_A \overline{w'q'} \right)$. Shallow mixed layers can quickly cool. A 10 meter mixed layer with a 100 W/m² of net upward heat flux will freeze (lake temperature somewhere below ~90 K) in a handful of Titan days (known as a tsol, which is equivalent to ~15.9 Earth days) while a 500 meter mixed layer with 100 W/m² of net upward heat flux requires a good part of a Titan season to do the same. Note that it is the lake mixed-layer depth, not the actual depth of the lake, the is relevant.

The saturation vapor pressure of liquid methane is calculated from the relation provided by Moses et al. [1992]. Binary ($N_2$ and $CH_4$) or more complicated solutions of liquids (e.g., addition of $C_2H_6$) are not directly accounted for, but could be approximated through a simple scaling constant that adjusts the saturation vapor pressure based on the assumed mixture of compounds. With respect to the moisture flux (Eq. 1), such a constant is effectively the same as altering the atmospheric humidity so that the difference between the atmospheric vapor content and saturation vapor content is appropriately changed.

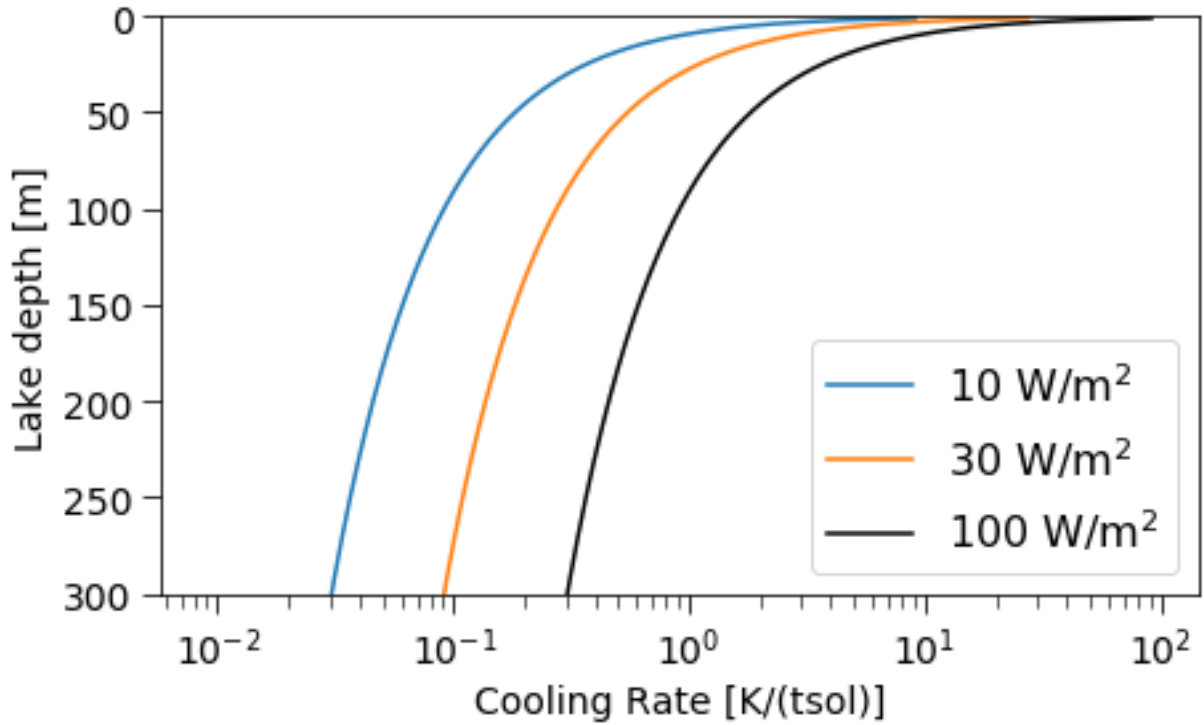

*Figure 3. Lake mixed layer cooling rates for a range of mixed layer depths and upward surface energy fluxes in the single slab lake model. The cooling rates are shown for a Titan day (tsol), which is ~15.95 Earth days.*

Over land, the surface is assumed to be dry and the surface temperature is fixed at the value specified at initialization. Latent heat fluxes are not calculated over land (i.e., no evaporation is permitted over land), but sensible heat fluxes are, and the constant land surface temperature implies that the ground serves as an infinite heat reservoir. The use of a constant land temperature is introduced for simplicity. In reality, there should be a transfer of energy between the land surface and subsurface, and this might tend to mute any exchanges with the atmosphere. For example, a turbulent cooling of the surface would like result in upward conduction of heat from the subsurface. If evaporation of a moist surface were allowed, a similar counter-response might be encountered. Future work will likely include an active regolith model and allow for evaporation from moist ground.

To initialize the mesoscale simulations, we started with vertical temperature and methane mixing ratio data obtained from the Huygens Atmospheric Structure Instrument (HASI) and the Gas Chromatograph Mass Spectrometer (GCMS) on the Huygens lander probe [Fulchignoni et al., 2005; Niemann et al., 2005]. This baseline sounding was then modified to meet the specific goals of the numerical modeling experiments. The resulting profiles were used to uniformly initialize the model. Because Titan's atmosphere exhibits only a small variation of temperature as a function latitude and longitude (several K from pole to pole [Jennings et al., 2011]), the HASI temperature profile is very similar to temperature profiles measured by the Cassini Radio Occultation experiment and the Voyager mission [Schinder et al., 2011], particularly in the lower portion of the atmosphere. Many of the radio occultation measurements were made at lake latitudes. Thus, using the HASI temperature profile in our soundings was reasonable for our idealized experiments.

For the initialization of methane humidity in the atmosphere, most simulations were started with a sounding that has zero atmospheric methane. To understand the effects of an initial non-zero atmospheric methane profile, two types of initial methane profiles were created (Fig. 4). The first profile type has a constant relative humidity from the surface up to 30 km, after which the profile follows a constant mixing ratio. Generally speaking, because temperature decreases with height, a constant relative humidity profile translates to a decrease in mixing ratio as a function of height. We classify this first type of profile as "unstable", because the initial virtual temperature profiles have vertical buoyancy due to the gradient in methane mixing ratio (Fig. 5). Virtual temperature takes into account the change in the ideal gas constant due to changes in atmospheric composition, and allows for the use of a single gas constant value ($R_d$, Table 2): $R_d T_v = RT$, where $T_v$ is the virtual temperature, $T$ is the temperature, and $R$ is the true gas constant based on the variable composition of the atmosphere. Since methane has a lower molar mass than nitrogen, adding methane will increase the virtual temperature. The virtual temperature and buoyancy effect are discussed in more detail below and in Section 3. The second profile type has a constant mixing ratio from the surface up to the height where the relative humidity reaches 100%. At this point, the mixing ratio follows the methane saturation curve. Since mixing ratio is constant in the lower atmosphere, and because temperature is decreasing with height, these profiles are statically stable; there is no initial buoyancy at the surface.

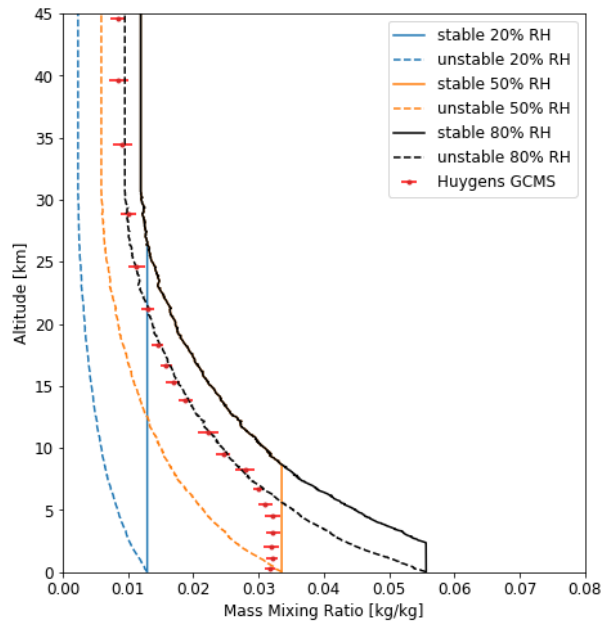

*Figure 4. The mixing ratio as a function of altitude for various methane soundings (lines), compared to the Huygens GCMS data (red points with error bars) taken from Niemann et al., [2010].*

For each type of methane profile, we generated profiles with an initial surface relative humidity of 20%, 50%, and 80%. Fig. 4 displays all six methane profiles compared to the methane profile measured by the Huygens GCMS [Niemann et al., 2010]. All three of the stable profiles eventually follow a 100% relative humidity curve at altitude. Fig. 5 shows the virtual potential temperature profiles calculated using the methane soundings in Fig. 4 and the temperature profile provided by the HASI data. Any profile where the virtual potential temperature decreases with height is statically unstable. The stable profile that starts with 50% relative humidity at the surface most closely resembles the Huygens GCMS data. By running simulations with both the unstable and stable methane profiles, we explore a wide range of possible methane profiles that may occur on Titan.

Most of the simulations were initialized with a profile that had no initial horizontal or vertical wind. In the few simulations that had an initial wind, a vertically uniform wind was applied to the sounding (see Table 2 for a list of simulations and their initializations).

The two-dimensional model domain is 1600 km wide with an approximately 20 km deep atmosphere. The width and depth were chosen so that lateral and top boundaries would be significantly distant from the lake, which is centered in the horizontal domain. The maximum lake dimension varies from 32 km to 300 km depending on the simulation. Horizontal grid spacing is 2 km, which is adequate for resolving the structure of a sea breeze. The lowest level vertical spacing is approximately 3 m and is gradually stretched with height to approximately 500 m spacing at the top of the model. There are 60 vertical levels.

The use of a 2-D vs. 3-D domain necessarily results in the inability to simulate some aspects of a realistic Titan scenario. From a purely dynamical perspective, vorticity stretching is not possible in 2-D, so the spin-up of a vertical vortex cannot be simulated. While such processes are of interest generally, they are far beyond the scope of this paper in which we are trying to understand the fundamentals of Titan's air-sea interaction. Real lakes also have irregular coastlines that can focus or defocus sea breeze or land breeze circulations, which can in turn lead to preferred areas of rising motion that cannot be simulated in 2-D. Nevertheless, there is a long tradition of using 2-D simulations to study terrestrial sea breezes and to extract the salient, underlying dynamics of circulations resulting from lakes and seas. We follow that tradition with the implicit understanding that 2-D simulations have limitations. Future work will likely move toward understanding how the additional third dimension modulates the results found in 2-D.

The horizontal boundary conditions of a mesoscale model (or any model) have the potential to generate non-physical signals that propagate through the rest of the model domain. This propagation of boundary-generated noise is particularly challenging for idealized simulations that require numerical conditions rather than physically-generated values that might come from a general circulation model. We experimented with open and periodic boundary conditions (Skamarock et al., 2008), as well as a number of different domain sizes in order to find an initial model configuration that generated numerically stable simulations over the longest duration possible. Open boundary conditions, sometimes called a radiation condition, allow waves, momentum, energy, etc. to leave the model domain and pass through the spatial boundary undisturbed. Ideally, an open boundary condition will not create reflections nor transmit information back into the model domain. A periodic boundary condition, avoids the issues with open lateral boundaries, but if the domain is too small, however, then energy, momentum, waves, etc. can wrap back into the region of interest and the system interacts with itself.. Using a large model domain helps mitigate the effects of a periodic domain. We desired to run these mesoscale simulations as long as reasonably possible to ensure that our experiments were reaching a dynamic and thermodynamic steady-state. To do so, we needed to minimize numerical noise generated by the boundary conditions. Based on numerous experiments, the mtWRF simulations were run with open or periodic boundary conditions that were set many 100s of km away from the edge of a simulated lake, which typically provides stable simulations that ran for at least 10 tsols. The numerical considerations of a very weakly forced system like Titan are of great enough concern that a section of this paper is dedicated specifically to this topic.

The base version of WRF used for the mtWRF model uses single precision for floating-point arithmetic. Due to the weak forcing in the Titan atmospheric system we found that the numerical precision caused inaccuracies in the lake surface temperature calculations. The lake surface temperature tendency was often so small that when multiplied by the timestep (1 s) and added to the prior total lake temperature, the single precision floating point variable was unable accurately represent the new, updated lake temperature. To account for this, we implemented a simple scheme that accumulated the temperature tendencies until the associated temperature change was large enough to be added to the previous lake temperature without loss to numerical precision. Once the accumulated lake temperature change was added to the lake temperature tendencies, the accumulated amount was zeroed and the accumulation process started over. The number of timesteps required to accumulate a large enough temperature change varied by grid point depending on the details of the energy exchanges at the lake surface, but was on the order of tens of timesteps (~10 s). Implementing this accumulation scheme for the lake surface temperatures had a noticeable impact on the evolution of the atmosphere and lake, and substantially improved energy conservation in the combined atmosphere-lake system. We checked other model physical processes for similar precision issues and did not find any obvious issues.

The model integration produces tendencies in SI units of seconds (s); however, units of hours or days in this paper are with respect to Titan. As previously indicated, a Titan day is known as a Titan sol or tsol for short, and there are 24 Titan hours in a tsol. A tsol is ~15.9 Earth days. Seconds retain their SI definition.

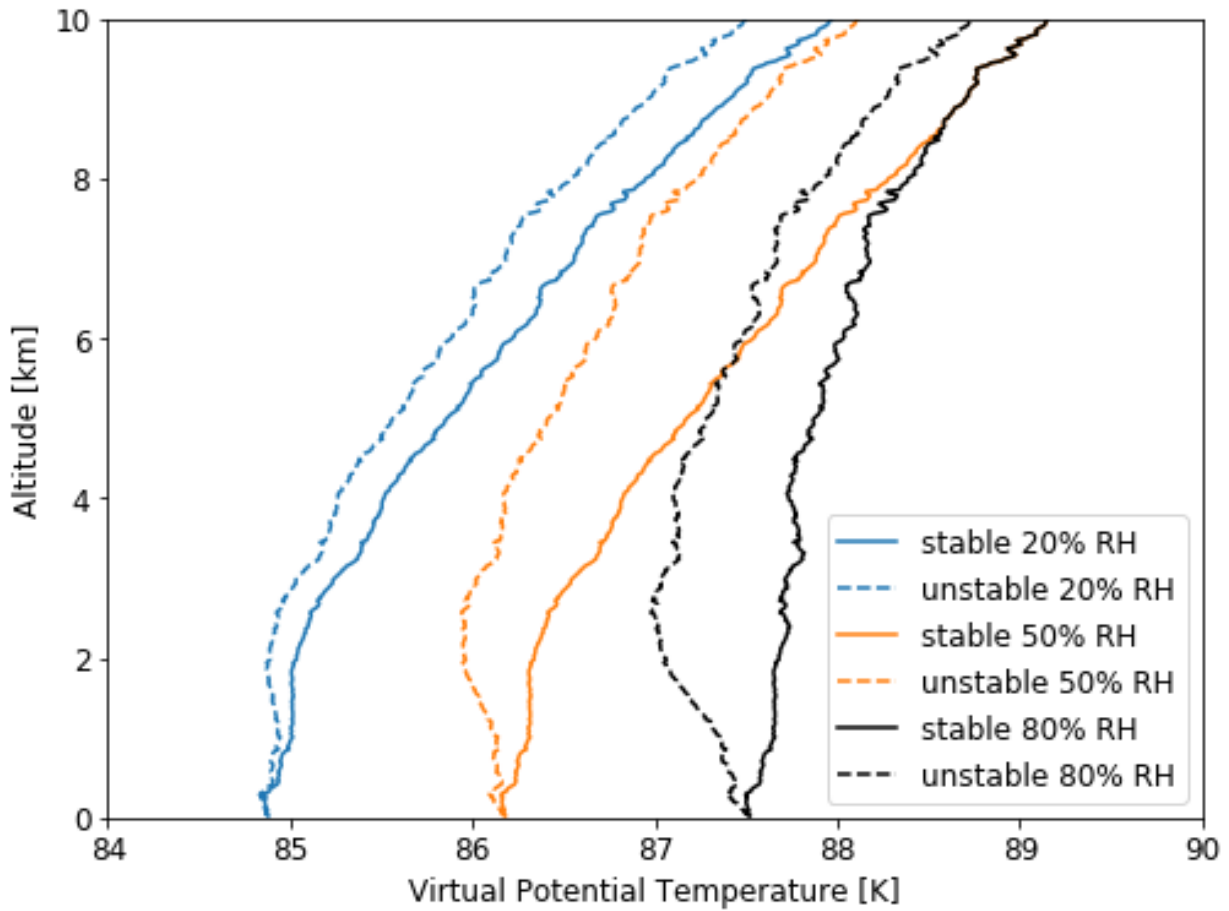

*Figure 5. The virtual potential temperature profiles of the six methane soundings, calculated by using the temperature profile from the HASI instument.*

## 3. The Canonical Circulation

All the simulations performed in this study show the same general solution; the different, specific solutions are variations on a theme that is illustrated schematically in Fig. 6. The canonical simulation is selected from the numerous collection of simulations, because it embodies many of the features and processes that appear in the other results. As such, it is an illustrative and instructive starting point to which other simulations can be compared and contrasted. At the same time, this canonical simulation does not necessarily present the most realistic simulation. From a practical standpoint, the canonical simulation was the first 300 km lake simulation we conducted, and all subsequent 300 km simulations flowed from this case. As the atmosphere is initially set to a subsaturated condition, there is an initial burst of evaporation. The increase in methane vapor produces a virtual temperature effect—an increase in positive buoyancy due to the lower molecular weight of methane compared to the moist nitrogen atmosphere. Compared to water vapor on Earth, methane is 18 g mol$^{-1}$/16 g mol$^{-1}$ = 112.5% more buoyant on a molecule per molecule basis. That buoyancy effect can be further magnified on Titan, because the saturation mixing ratio of methane is usually much higher than water. The result of the positive buoyancy is a rising plume of moist air and the accompanying circulation demanded by mass conservation: A land breeze converging over the center of the lake and a divergent circulation aloft. The circulation is generally fully mature in less than 6 hours, although there is some variability in the development time depending on the specifics of the modeling scenario.

The evaporating lake cools as the land breeze intensifies, which initiates a sensible heat transfer from the atmosphere to the lake and the cooling of the near-surface atmosphere. Within an hour or two, the growing pool of cold air over the lake initiates a direct thermal circulation—a sea breeze—circulating in the opposite sense to the land breeze. Generally, the sea breeze circulation tends to be shallower than the land breeze, and destructive interference is most pronounced in the lowest layers of the atmosphere. The result is a decrease of surface winds above the lake. The decrease in wind speed in conjunction with the increasingly stable atmosphere leads to a diminishment of all fluxes (Eq. 1). The final quasi-steady state is a very shallow, cold, stable, and moist (but not saturated) marine boundary layer over the lake with a dominant sea breeze structure in the low levels and a weak background land breeze circulation most evident aloft, above the sea breeze circulation, and extending farther out over the land. We occasionally refer to the land breeze circulation as a plume circulation because is it represents a rising plume of methane-enhanced air with inflow at the low levels and outflow aloft. Fluxes over the lake are small due to low wind speeds and increased stability. The magnitude of the latent heat flux is also reduced, because the cooling lake lowers the saturation vapor pressure over the lake, and when combined with the moistening atmosphere, the vertical moisture gradient between the atmosphere and lake is reduced (Eq. 1). The final, quasi-steady solution is a classical cool and stable marine boundary layer.

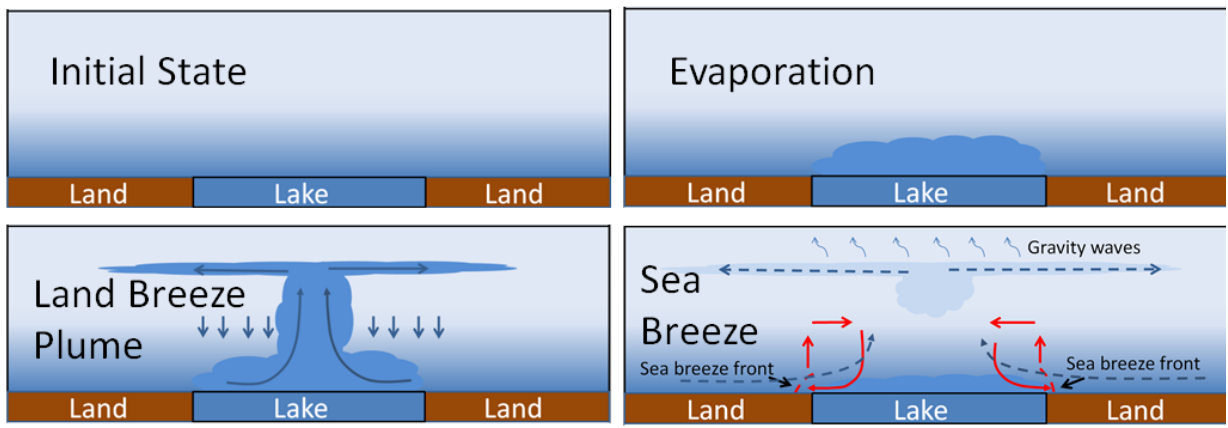

*Figure 6. Idealized evolutionary states of an atmospheric circulation associated with a Titan lake. Note that blue shading in the atmosphere represents vapor structures not cloud. Starting from an initial static state with a subsaturated atmosphere and identical lake and atmosphere temperatures (top left), evaporation results in the moistening of the atmosphere directly above the lake (top right). The moist air is buoyant, and a plume of methane-rich air rises, which establishes a land breeze circulation (dark arrows, bottom left). The lake cools due to evaporation, and the atmosphere above the lake begins to cool through sensible heat fluxes. As the air cools, density increases and begins to drive a sea breeze circulation (red arrows, bottom right) acting in opposition to the land breeze (dashed arrows, bottom right). Usually, the sea breeze overwhelms the buoyancy circulation, but a weak, residual plume*

*circulation remains and extends beyond the lake domain. The sea breeze circulation is generally confined to the lake. The cooling air increases atmospheric stability, which tends to decrease latent and sensible heat fluxes. The decreasing fluxes and increasing stability limit the sea breeze strength. A sea breeze front demarcates a narrow transition between the land air mass and the marine layer. Gravity waves emanate from the circulations.*

Of course, on real Titan the atmosphere doesn't "spin-up" from some arbitrary initial condition. The model spin-up period might be regarded as an unrealistic state to be neglected until a more steady-state solution is obtained. Complete neglect of the spin up solution, however, would overlook the tendency for a real buoyancy circulation to operate behind whatever net, steady-state circulation may be present. The forcing behind a buoyant land breeze circulation is a real physical process. The strength of the initial plume circulation, the strength of the sea breeze, and the overall evolution and balance between these two opposing circulations depends on a variety of factors including the temperature differential between the lake and that atmosphere, the initial atmospheric humidity, and the depth of the lake mixed layer.

Results from a simulation with a 300 km-wide lake illustrate well the canonical solution (Fig. 7). For reference, Ligeia Mare, the second largest lake on Titan has dimensions of approximately 350 x 420 km. The largest lake, Kraken Mare is closer to 1000 km x 400 km. Thus, a 300 km lake (in 2-D) is representative of the larger lakes on Titan. The fields are averaged over a period of two Titan hours in order to mute transient (but physically real) circulations that would otherwise obfuscate the bulk circulation. The vertical domain is truncated in the plots so that greater detail can be seen in the lower atmosphere where the bulk of the circulation is found. Recall that the top of the model domain is approximately 20 km.

The initial plume circulation is evident within the first 2-hour averaging period (Fig. 7a). A collection of small scale, methane-rich plumes over the lake have risen to a level near 1 km. Inflow toward the lake is produced at the lake shore below 1 km and outflow is found above, up to an altitude of ~2 km. The collective width of the enhanced vapor is approximately the same as that of the lake, where the effective width of the plume circulation is qualitatively defined as the leading edge of the upper level outflow (i.e., where the wind changes from a substantive non-zero wind speed value in the outflow to a near zero unperturbed environmental wind). The zero mixing ratio contour line indicates that a limited amount of moisture has made its way inland over the short period of time. The circulation accompanying the buoyant plume is clearly defined by low level inflow and upper level outflow. Over time, this circulation would be expected to keep the low-level moisture horizontally confined close to the lake while advecting methane to greater distances in the outflow aloft. The high methane mixing ratios in the shallow atmospheric layer just above the lake will eventually evolve into the stable marine boundary layer. The peak average wind speeds are approximately 0.5 m/s, although the surface winds are smaller in magnitude due primarily to the effect of surface friction.

After one tsol (Fig. 7b), a sea breeze circulation has been established. It is confined roughly to the dimensions of the lake and is, initially, approximately the same depth (~3 km) as the plume circulation, which has continued to expand vertically and horizontally. Note that the sea breeze circulation completely dominates the plume circulation over the lake.

At two tsols (Fig. 7c), the sea breeze and background plume circulation are fully developed. The sea breeze circulation remains confined to the lake vicinity while the background plume circulation extends far beyond the lake boundary. The transition between the low level land breeze from the plume and the sea breeze forms a sharp sea breeze front. There is enhanced export of vapor at altitude while the low-level vapor remains largely confined within the lake boundaries. Thus, it appears that the juxtaposition of the two circulations play a role in how methane is transported outside the local lake region.

At three tsols (Fig. 7d) the solution is very similar to the solution at two sols; a steady-state solution has been obtained. Over the lake, a mature sea breeze circulation is present, and the marine boundary layer remains confined to the immediate vicinity of the lake. Any vapor found at a distance from the lake is due almost entirely to the transport from the plume circulation. Above the main sea breeze circulation cell is the plume circulation outflow, and above that are vertically damped gravity waves (not shown) that reflect alternating patterns of wind with a vertical wavelength similar to that of the plume circulation below. The overall vertical scale of the combined sea breeze and larger buoyancy circulation is at ~3.5 km. The sea breeze circulation is confined to the lowest ~1.5 km and the marine layer is only 200 m to 300 m deep over the lake with a rise in height at the sea breeze front.

Fig. 8 focuses on the rightmost half of the sea breeze. The structure of the leftmost sea breeze is nearly identical. Three distinct regions can be identified based on wind, vapor, and temperature structure. The first region is the marine layer directly over the lake. The second region is a transition region extending from the lakeshore (x=950 km) to the sea breeze front located at approximately x=1000 km. The third region is the nearly unmodified atmosphere further inland. The sea breeze does continue to move slowly inland with time even though the structure is quasi-steady state. By five tsols, the sea breeze front is at x=~1100 km (not shown). The frontal displacement of ~100 km over two tsols gives a propagation speed of ~3 cm/s.

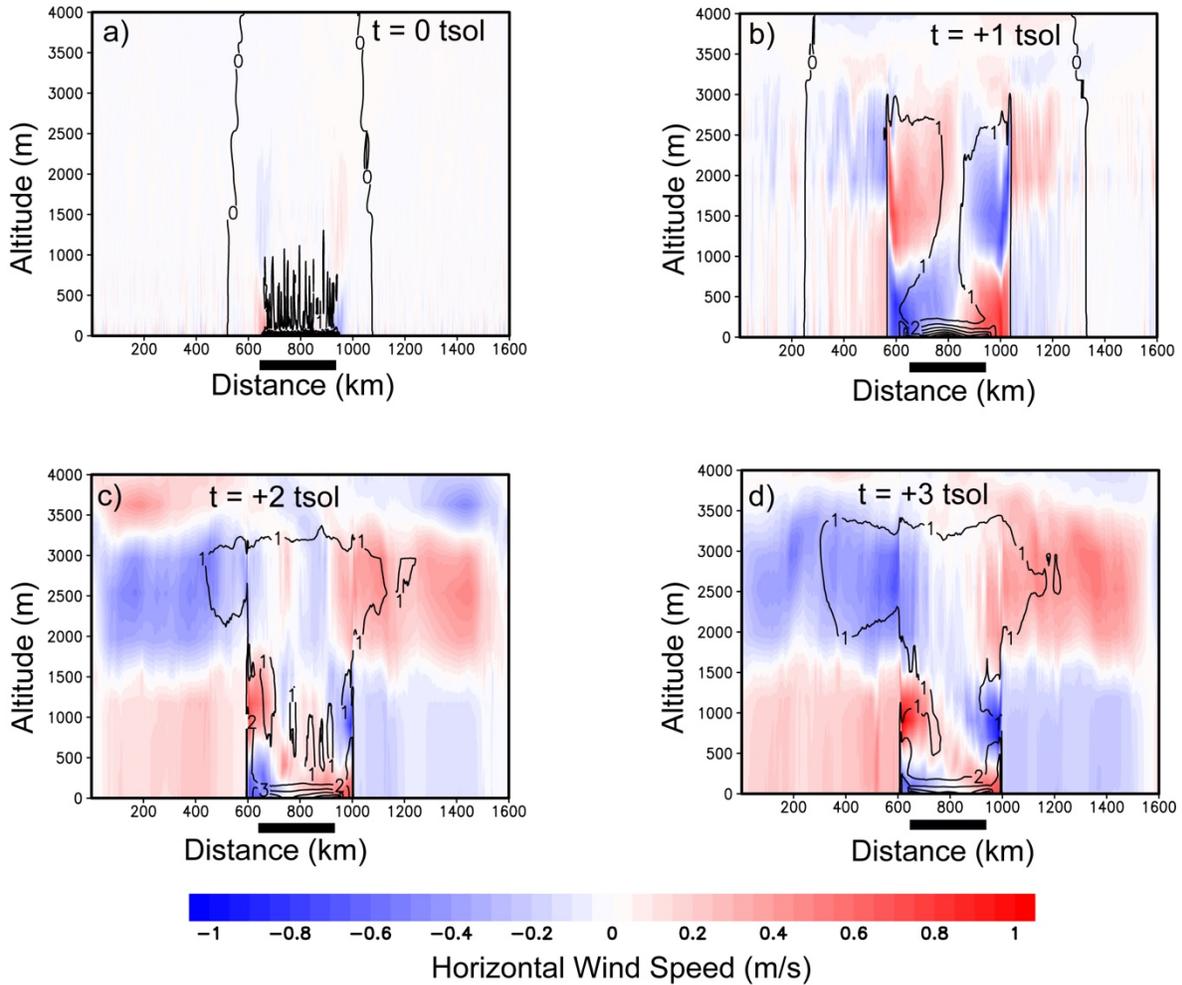

*Figure 7. Bulk sea breeze evolution for a 300 km lake. Horizonal velocity is shaded with red (positive) values indicating flow from left to right and blue (negative) values indicating flow from right to left. Contours are methane mixing ratio (g/kg). The 300 km lake domain is shown by the black bar below the horizontal axis. The portion of the initial buoyant vapor plume over the lake (a) is quickly replaced by a quasi-steady sea breeze circulation at low levels with fronts that remain just beyond the coast in (c) and (d), but the broader plume circulation remains in the background. All fields are averaged over a two hour period with the referenced time indicating the starting time of the averaging period.*

The marine region is characterized by an extremely stable and shallow (< 300 m) boundary layer with a relative humidity that peaks at only ~20%. This is moist compared to the initial dry atmosphere, but it also demonstrates how inefficient evaporation is at substantially moistening the boundary layer. The temperature in the lowest atmospheric layer is 90.9 K compared to almost 94 K farther inland. The layer is not well mixed, and vertical motion is suppressed in the highly stable layer. The onshore flow in the marine layer increases as a function of height up to the top of the temperature inversion. The onshore flow persists up to ~400 m where it reverses to form the sea breeze return flow. Vertical motion is more evident above the sea breeze circulation where the atmospheric stability is lower and where weak upward motion is associated with the plume circulation. Although the marine layer is very shallow, the atmosphere over the lake is colder than the atmosphere over land up to an altitude of ~1 km. The marine layer is moist, but it is not near saturation, and no clouds would be expected to form. A more detailed analysis of the possibility of clouds is given in Section 8.

The transition region is marked by a rapid decrease in sea breeze flow and the appearance of increasingly negative vertical velocity in the low levels. The vapor field shows an increase in the atmospheric mixed layer depth from the nominal ~200 m at the shoreline up to ~800 m at the front, although the highest values of moisture remain over the lake. Temperatures are notably warmer at the lakefront and increase moving inland towards the front. The marine inversion gradually erodes until it is completely gone at the front.

The sea breeze front is marked by a sharp transition from the moist marine layer into the very dry air mass over land. What vapor there is over land was predominantly transported there by the upper portion of the plume circulation. There is also a well-defined increase in temperature horizontally across the front up to a height of ~1 km. Downward velocity peaks at the front, at nearly 25 cm/s within 100 m of the surface. Vertical velocity remains negative on the land side of the sea breeze in association with the descending branch of the plume circulation. By Earth standards, both the horizontal and vertical wind speeds are small, but on Titan, these perturbations are modest compared to Titan's putatively sluggish large-scale circulation and may very well be typical in lake regions.

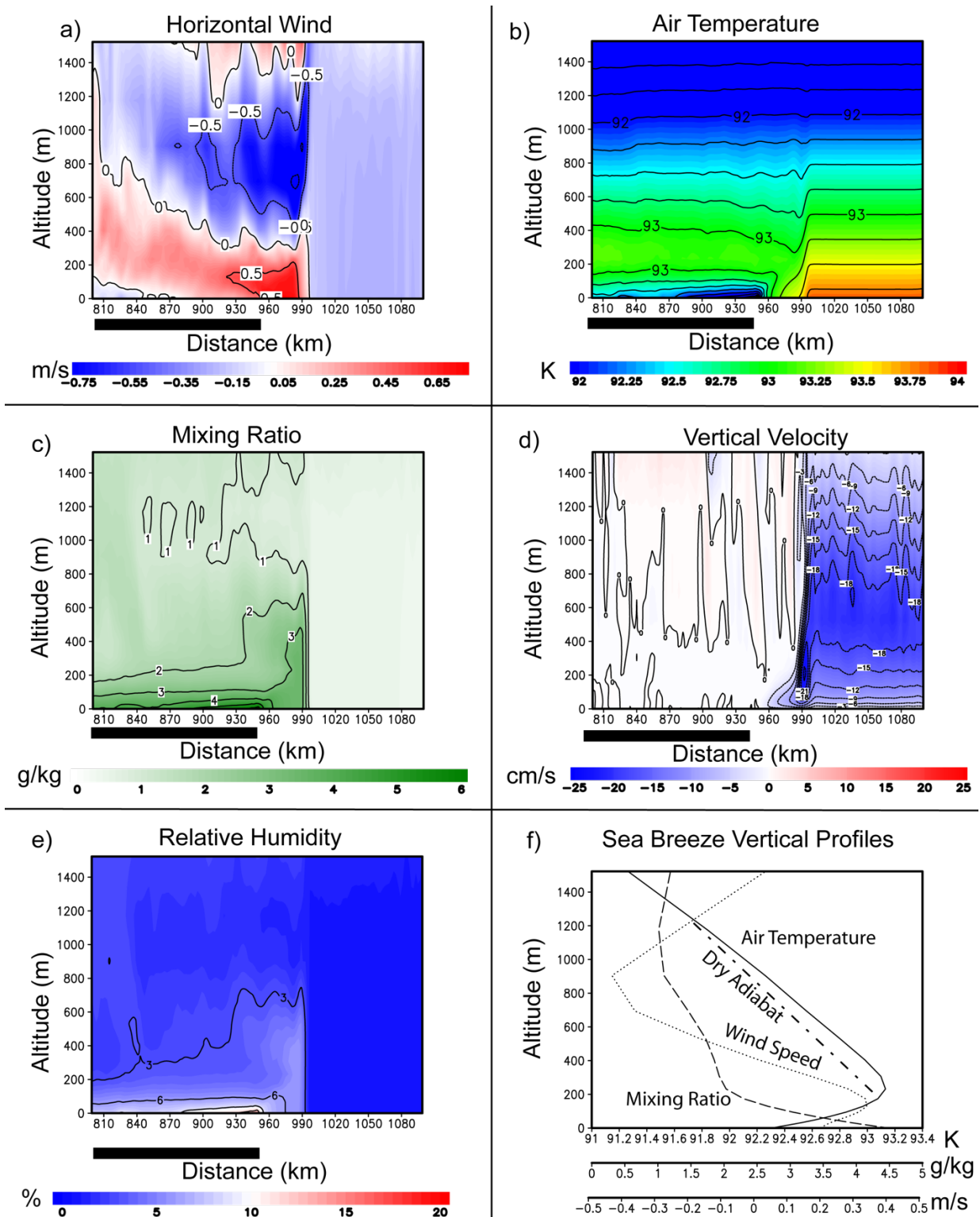

*Figure 8. The instantaneous (not averaged) sea breeze and marine boundary layer structure (a-e) for Simulation 4 at 3 tsols. The domain occupied by the lake is shown as a black bar on the horizontal axis. Panel (f) is also instantaneous in time, but horizontally averaged from 800 km to 1000 km.*

Analysis of the lake temperature, near-surface wind and air temperature, and surface fluxes explains the behavior and structure of the atmosphere over the three regions. Hovmöller diagrams (Fig. 9) provide a good overview of the variations and their relationship to one another. Additional insight is gained by spatially averaging the properties over the lake to a single value and exploring the variations over time (Fig. 10). The time series of properties at a single near-surface point (Fig. 11) is also instructive, because it brings out details that are lost in the averaging and which are difficult to visually extract from the Hovmöller presentations.

The effect of the initial plume circulation is evident in Fig. 9, but may be difficult to see, because the signal of the plume is short-lived in the earliest part of the simulation. A burst of latent heat flux occurs at the start of the simulation. In association with the latent heat flux the lake begins to cool, and a non-zero sensible heat flux follows. In a few hours or less, the latent heat fluxes over the lake drop back down to smaller values due to the establishment of the stable marine boundary layer. In contrast, the sensible heat flux at the shoreline and just inland are at a maximum once the marine layer develops. This is due to the cold marine air flowing over the land within the sea breeze coupled with the ground temperature being held constant.

Regardless of the magnitude of latent and sensible heat flux over the lake, the Bowen ratio field is remarkably smooth without any fine-scale spatiotemporal structure. The Bowen ratio is defined as the ratio of the sensible heat flux to the latent heat flux. The ratio slowly increases in magnitude in time (becoming increasingly negative), eventually approaching -1.0 near the shoreline. Thus, even though there is temporal variation of the sensible and latent heat fluxes, they act in unison to smooth out the time variations in the Bowen ratio. This should not be altogether surprising since both fluxes are computed using the same bulk aerodynamic formulation (1) with identical density, winds, and aerodynamic coefficient. The lake surface temperature shows a similar pattern suggesting that it is the dominant controlling factor in the Bowen ratio. The constancy of the Bowen ratio is found not only in the spatial average (Fig. 10) but also instantaneously at any given point (Fig. 11). There is a very strong coupling between the sensible and latent heat flux such that their ratio is nearly invariant no matter the magnitude of turbulent flux activity.

The Hovmöller diagrams (Fig. 9) show strongly coherent spatial structures except in the Bowen ratio and lake surface temperature that have weaker coherent structures. Some of these structures move from the center or even from one side of the lake toward the opposite coast. As these patterns occur simultaneously (i.e., the patterns cross), at least some of the features propagate counter to the prevailing wind and cannot be purely advective. Gravity waves launched at the sea breeze front and those associated with the plume circulation are likely genetic candidates. As the coherent structures move across the lake, the fluxes respond to the small changes in wind speed, temperature and vapor. Where the winds are strongest, the fluxes are largest, and changes in air temperature and vapor follow from that so that the coherent structures are consistent and visible through all these fields. The turbulent exchanges are not uniform in space or time and may be characterized as "bursty".

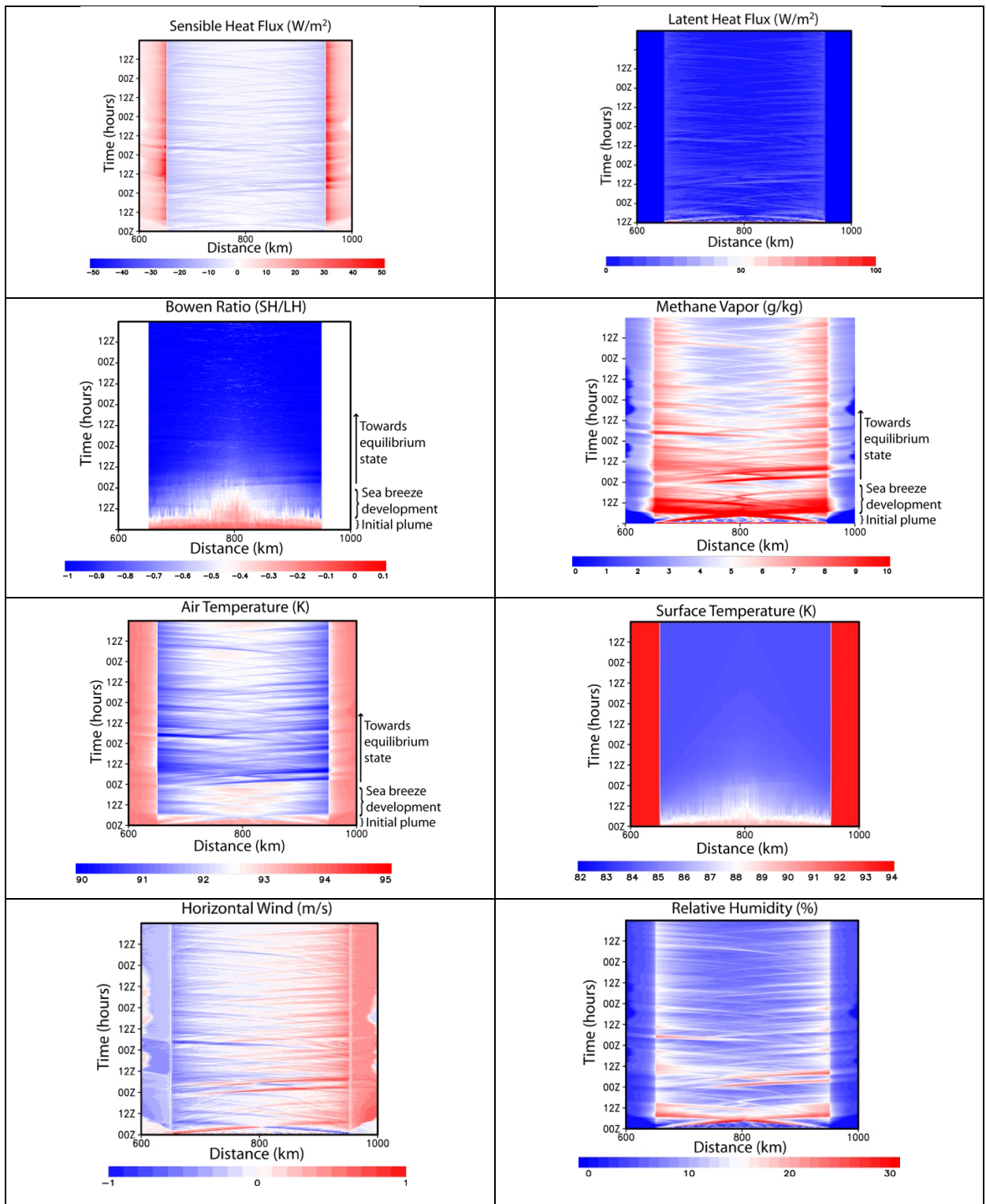

*Figure 9. Hovmoller diagrams of canonical lake and atmospheric properties from Sim4. The spin-up period dominated by a vapor plume circulation is only a few hours. After spin-up, a sea breeze circulation develops superimposed on the plume circulation. The different stages of development form plume to sea breeze to equilibrium solution is identified in some of the panels. The lake extends from 650 to 950 km.*

For the first two tsols, the air-sea interaction activity appears stronger than in the later tsols. By the last sol of the simulation, the fields have generally stabilized. Methane vapor and relative humidity are highest near the coastline. The air generally flows away from the center of the lake toward the coastlines, and along that trajectory the air continually cools and moistens. Even so, maximum relative humidity over the lake is no more than 20%. The air accelerates as it moves across the lake with the strongest winds found along the coast. The winds are not steady, however. Peak winds at the coast can sporadically approach 1 m/s. Winds are weaker farther over the lake, and outside of the sporadic bursts of activity, the wind is often nearly calm.

On average (Fig. 10), the air over the lake cools slightly, while the temperature of the lake drops close to 84 K. In most reasonable compositional scenarios, the lake would freeze before reaching this temperature. Assuming the lakes on Titan do not regularly freeze, this result suggests that the specific initial conditions are not entirely realistic, or perhaps important physics are still missing. Both of these points are addressed through the numerous additional simulations described later.

Once the quasi-equilibrium solution is reached, the average fluxes over the lake (Fig. 10) settle to around 5 W/m$^2$ or less. Yet, looking at any given time in Fig. 11, it is clear that there is considerable variation around this average. The time variability of the air-sea exchange process is masked by the average, as the bulk of the exchange occurs in bursts where the fluxes rise up to 15 W/m$^2$ in magnitude before dropping down to values closer to zero during more quiescent periods. Likewise, the wind speed is very close to zero, except during the sporadic bursts of activity when it rises to ~20 cm/s. The instantaneous values (Fig.

11) have a signature of the coherent structures seen in Fig. 9, as they pass over a point on the lake. These periods are well correlated with an increase in fluxes, wind speed, mixing ratio, and relative humidity. The change in humidity is due both to changes in vapor content and air temperature; vapor content is usually found to go up when temperature goes down. The anti-correlation of temperature and vapor is consistent with a moistening due to evaporation and a near simultaneous cooling of the air through a sensible heat flux with the cooling lake.

The Bowen ratio does equilibrate to near -1.0, as proposed by M07. A more detailed look at the model solutions compared to the M07 analytical results is presented in Section 5. Although the equilibrium Bowen ratios are -1.0 in both the model and the analytical results, the fluxes and atmosphere and lake properties are not necessarily identical to M07 owing to the representation of atmospheric dynamics and thermodynamics in the mtWRF simulation.

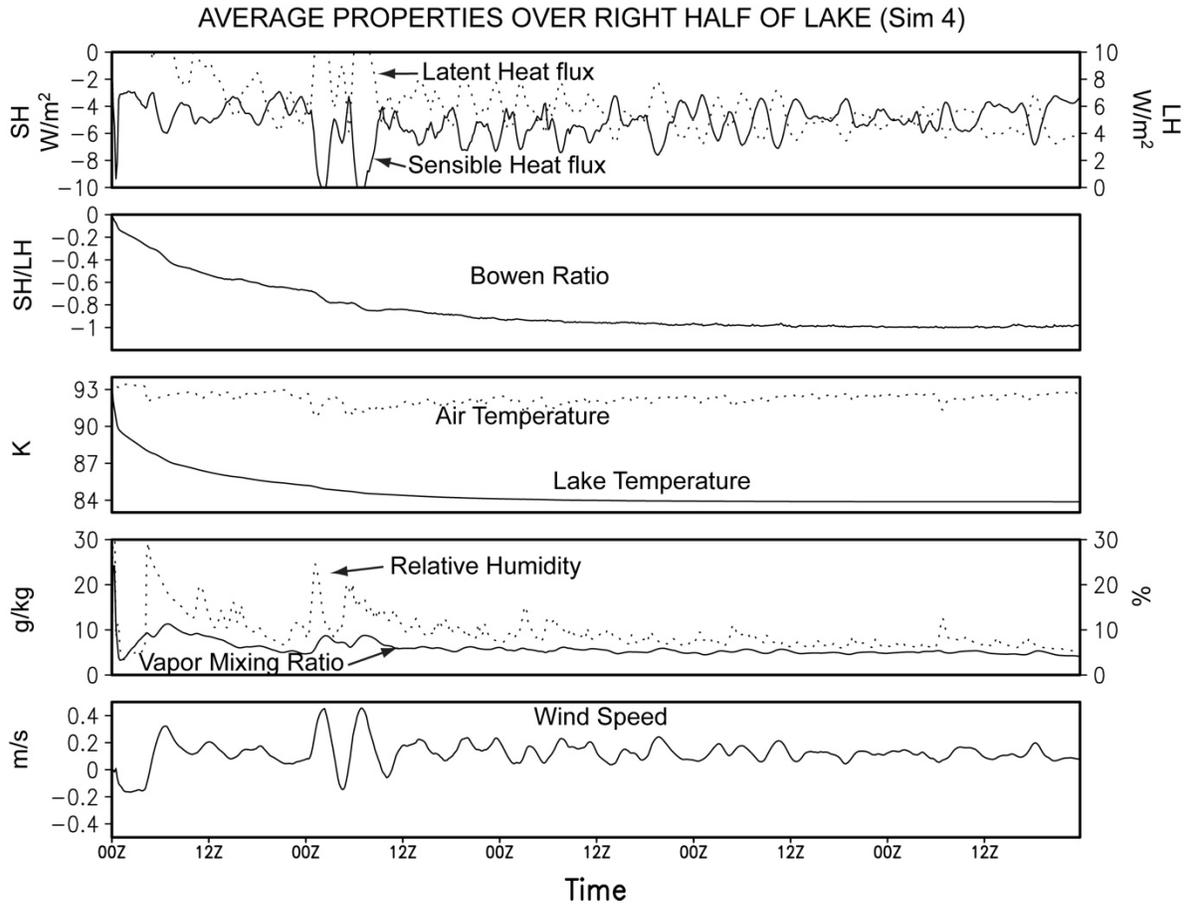

*Figure 10. Average properties over right half of the lake for Sim4 (300 km lake). Top to bottom: Sensible heat flux (solid W/m$^2$) and Latent heat flux (dotted W/m$^2$); Bowen ratio; SST (solid K) and air temperature (dotted K); methane mixing ratio (solid g/kg) and relative humidity (dashed %); wind speed (m/s).*

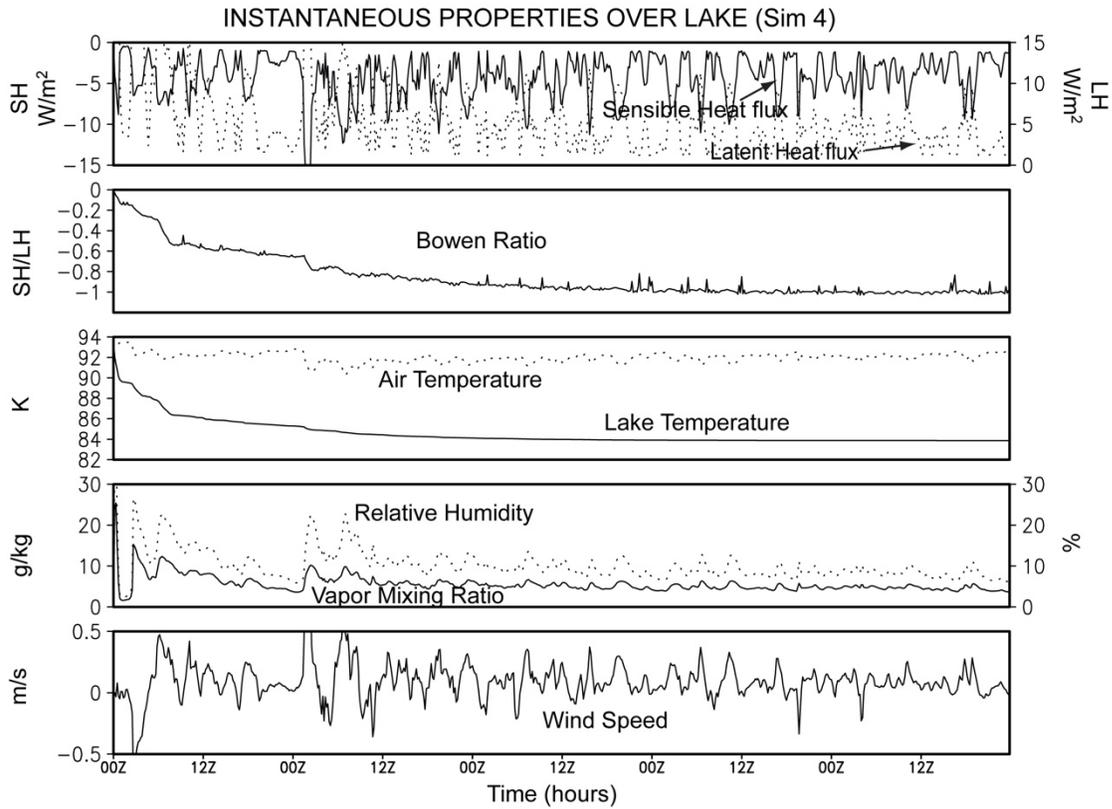

*Figure 11. Instantaneous values 75 km from right shoreline of the lake for Sim4 (300 km lake). Top to bottom: Sensible heat flux (solid W/m²) and Latent heat flux (dotted W/m²); Bowen ratio; SST (solid K) and air temperature (dotted K); methane mixing ratio (solid g/kg) and relative humidity (dashed %); wind speed (m/s).*

## 4. Parameteric Studies

Variations from the canonical solution are closely tied to a specific set of initial parameters. Changing these parameters produces variations in the evolution of the system, which we investigated. The variable parameters include lake size, lake mixing depth, initial lake and atmosphere temperatures, atmospheric relative humidity, and mean background wind (Table 2). We also experimented with different subgrid scale diffusion settings. While changes to diffusion slightly altered the solution, it did not change any important details. Since mesoscale simulations are potentially sensitive to the choice of boundary conditions, we ran experiments with the three different boundary conditions built into the WRF model: open, periodic, and symmetric (see Skamarock et al., 2008, for details; note that symmetric here refers to the boundary condition and not the configuration of the domain). The canonical solution in previous figures was taken from Simulation #4.

Table 2. Key parameter settings for the mtWRF simulations

| Simulation | Lake Size (km) | Mixed Layer Depth (m) | Lake Temp (K) | Land Temp (K) | Initial Wind (m/s) | Relative Humidity (%) | Boundary Condition | Duration (tsols) | Notes |
|---|---|---|---|---|---|---|---|---|---|
| 1 | 32 | 1 | 93.47 | 93.47 | 0 | 0 | open | 10 | Initial test case. Shallowest mixed layer. |
| 2 | 32 | 1 | 93.47 | 93.47 | 0 | 0 | periodic | 10 | Same as #1, but with different b.c. |
| 3 | 100 | 1 | 93.47 | 93.47 | 0 | 0 | periodic | 5 | Same as #2, but with wider lake. |
| 4 | 300 | 1 | 93.47 | 93.47 | 0 | 0 | periodic | 5 | Canonical simulation, same as #1 and #2 but with 300 km lake. |
| 5 | 100 | 10 | 93.47 | 93.47 | 0 | 0 | periodic | 5 | Same as #3 but with 10 m mixed layer. |
| 6 | 100 | 100 | 93.47 | 93.47 | 0 | 0 | periodic | 5 | Same as #3 but with 100 m mixed layer. |
| 7 | 100 | 30 | 93.47 | 93.47 | 0 | 0 | periodic | 5 | Same as #3 but with 30 m mixed layer. |
| 8 | 100 | 100 | 93.47 | 93.47 | 1 | 0 | periodic | 5 | Same as #6 but with 1 m/s wind. |
| 9 | 100 | 100 | 93.47 | 93.47 | 3 | 0 | periodic | 5 | Same as #6 but with 3 m/s wind. |
| 10 | 100 | 100 | 94.47 | 93.47 | 0 | 0 | periodic | 5 | Same as #6 but with 1 K warmer lake. |
| 11 | 100 | 100 | 92.47 | 93.47 | 0 | 0 | periodic | 5 | Same as #6 but with 1 K colder lake. |
| 12 | 100 | 100 | 93.47 | 94.47 | 0 | 0 | periodic | 5 | Same as #6 but with 1 K warmer land. |
| 13 | 100 | 100 | 93.47 | 92.47 | 0 | 0 | periodic | 5 | Same as #6 but with 1 K colder land. |
| 22 | 100 | 1 | 93.47 | 93.47 | 0 | 0 | symmetric | 5 | Same as #3 but with symmetric b.c. |
| 23 | 300 | 10 | 93.47 | 93.47 | 0 | 0 | periodic | 5 | Canonical (#4) but with 10 m mixed layer. |
| 24 | 300 | 100 | 93.47 | 93.47 | 0 | 0 | periodic | 5 | Canonical (#4) but with 100 m mixed layer. |
| 25 | 300 | 30 | 93.47 | 93.47 | 0 | 0 | periodic | 5 | Canonical (#4) but with 30 m mixed layer. |
| 26 | 300 | 100 | 93.47 | 93.47 | 1 | 0 | periodic | 5 | Same as #24 but with 1 m/s wind. |
| 27 | 300 | 100 | 93.47 | 93.47 | 3 | 0 | periodic | 5 | Same as #24 but with 3 m/s wind. |
| 28 | 100 | 100 | 87.47 | 93.47 | 0 | 0 | periodic | 5 | Same as #6 but with 6 K cold lake. |
| 29 | 100 | 100 | 89.47 | 93.47 | 0 | 0 | periodic | 5 | Same as #6 but with 4 K cold lake. |
| 30 | 100 | 100 | 91.47 | 93.47 | 0 | 0 | periodic | 5 | Same as #6 but with 2 K cold lake. |
| 31 | 300 | 100 | 87.47 | 93.47 | 0 | 0 | periodic | 5 | Same as #24 but with 6 K cold lake. |
| 32 | 300 | 100 | 89.47 | 93.47 | 0 | 0 | periodic | 5 | Same as #24 but with 4 K cold lake. |
| 33 | 300 | 100 | 91.47 | 93.47 | 0 | 0 | periodic | 5 | Same as #24 but with 2 K cold lake. |
| 34 | 300 | 1 | 93.47 | 93.47 | 0 | 20% | periodic | 5 | Canonical (#4) but with "unstable" 20% constant RH. |
| 35 | 300 | 1 | 93.47 | 93.47 | 0 | 50% | periodic | 5 | Canonical (#4) but with "unstable" 50% constant RH. |
| 36 | 300 | 1 | 93.47 | 93.47 | 0 | 80% | periodic | 5 | Canonical (#4) but with "unstable" 80% constant RH. |
| 51 | 100 | 100 | 91.47 | 93.47 | 3 | 0 | periodic | 5 | Same as #30 but with 3 m/s wind. |
| 52 | 300 | 100 | 87.47 | 93.47 | 0 | 80% | periodic | 5 | Same as #31 but with "unstable" 80% constant RH. |
| 53 | 300 | 1 | 87.47 | 93.47 | 0 | 80% | periodic | 5 | Canonical (#4) but with 6 K cold lake and "unstable" 80% constant RH. |
| 54 | 300 | 1 | 87.47 | 93.47 | 0 | 0 | periodic | 5 | Canonical (#4) but with 6 K cold lake. |
| 55 | 300 | 1 | 89.47 | 93.47 | 0 | 0 | periodic | 5 | Canonical (#4) but with 4 K cold lake. |
| 56 | 300 | 100 | 93.47 | 93.47 | 1 | 0 | open | 5 | Same as #26 but with open b.c. |
| 57 | 300 | 100 | 93.47 | 93.47 | 3 | 0 | open | 5 | Same as #27 but with open b.c. |
| 58 | 100 | 100 | 91.47 | 93.47 | 3 | 0 | open | 5 | Same as #57 but with 100 km wide lake. |
| 59 | 300 | 1 | 91.47 | 93.47 | 0 | 0 | periodic | 5 | Canonical (#4) but with 2 K cold lake. |
| 60 | 300 | 1 | 93.47 | 93.47 | 0 | 20% | periodic | 5 | Canonical (#4) but with "stable" 20% RH methane profile. |
| 61 | 300 | 1 | 93.47 | 93.47 | 0 | 50% | periodic | 5 | Canonical (#4) but with "stable" 50% RH methane profile. |

| | | | | | | | | |
|---|---|---|---|---|---|---|---|---|
| 62 | 300 | 1 | 93.47 | 93.47 | 0 | 80% | periodic | 5 | Canonical (#4) but with "stable" 80% RH methane profile. |
| 63 | 300 | 100 | 87.47 | 93.47 | 0 | 80% | periodic | 5 | Same as #52 but with "stable" 80% RH methane profile and 6 K cold lake. |
| 64 | 300 | 1 | 87.47 | 93.47 | 0 | 80% | periodic | 5 | Same as #62 but with 6 K cold lake. |
| 65 | 300 | 1 | 93.47 | 93.47 | 1 | 0 | open | 5 | Canonical (#4) but with 1 m/s and open b.c. |
| 66 | 300 | 1 | 93.47 | 93.47 | 3 | 0 | open | 5 | Canonical (#4) but with 3 m/s and open b.c. |
| 67 | 100 | 1 | 91.47 | 93.47 | 3 | 0 | open | 5 | Same as #58 but with 1 m mixed layer. |

## 4.1. Variations in Lake Dimension

A 300 km lake is comparable in dimension to the largest lakes on Titan—Kraken Mare and Ligeia Mare. There are numerous smaller lakes that collectively could provide sources and sinks of heat and moisture similar to that of the largest reservoirs. It is worthwhile to test in the model whether smaller lakes drive similar circulations to the canonical solution.

Simulations #2 and #3 are identical to the baseline canonical case (Simulation #4) except for the lake size (32 km and 100 km wide lakes, respectively). Snapshots of the atmospheric circulation for these two simulations shortly after initialization and at tsol=3 are shown in Fig. 12, and may be compared with Fig. 7. Both simulations show an initial buoyant moisture plume in the first two hours. The width of the moisture plume scales with the size of the lake; the smaller the lake the narrower the plume. The depth of the plume, which has a dependence on the vapor excess over the lake compared to the land, also shows a dependence on the lake size. The 32 km lake plume circulation rises to ~2.5 km, the 100 km to ~3.0 km and the 300 km to ~3.5 km. The increasing depth may be most readily be explained by a greater entrainment of dry air by the narrower plumes. This is not dissimilar to entrainment effects on terrestrial clouds [e.g. Gregory, 2001; Redelsperger et al., 2002]. The depth of the inflow layer is comparable in all simulations (~1.5 km). The depth of the outflow layer is slightly larger for the larger lakes. The magnitude of the inflow and outflow are all comparable, but there is a trend for the larger lakes to have greater moisture in the outlow. This trend is consistent with lower entrainment and slightly greater buoyancy. The vertically propagating gravity waves are more evident at the top of the domain in the smaller lake simulations.

By one tsol, all the simulations show the development of a sea breeze superimposed on the preexisting and larger buoyancy circulation (not shown). As in the 300 km simulation, the sea breeze is confined very close to the coastline, but it does propagate slowly inland over time. A similar, very shallow marine layer forms in all the simulations. Fig. 13 shows the time series of average lake properties, which can be compared to Fig. 10. The trends in all simulations are very similar. The initial large spike in fluxes associated with the buoyant plume is present, followed by the relaxation to quasi-steady circulation with a superimposed sea breeze. The cooling of the lake to ~84 K, the cooling of the atmosphere, the march of the Bowen ratio toward ~-1.0 over the five tsol period, the very low wind speeds, and general decrease in sensible heat flux to increasingly small values is consistent between all the runs. The 300 km (sim #4), 100 km (sim #3), and 32 km (sim #2) lakes all have marine layer relative humidity of <~10%, but the larger the lake, the slightly larger the humidity. The likely explanation for this difference is fetch distance and the ability for the circulation to more easily mix drier air from land into the air over the lake. Regardless, the atmosphere does not continue to moisten despite continuous evaporation.

An unexpected but important result is that some lake simulations tend to become increasingly asymmetric with time. The asymmetry is a non-physical solution, because the simulation is constructed to be symmetric about the lake with symmetric forcing; whatever occurs on the left side of the domain should be exactly mirrored on the right side. Not all the simulations exhibit such a strong break from symmetry. Once asymmetries are established, they can be exacerbated by the lateral boundary conditions. The reason and implications of the breakdown of symmetry is discussed later (See Section 9).   As an example, in the 100 km simulation (Sim #3) the overall lake properties and evolution of fluxes are in family with the overwhelming trends and tendencies from the entire ensemble of parametric studies with stable and symmetric solutions, but the transient effect of the asymmetry is noticeable in the flux perturbations after tsol 3 (Fig. 13). So, while the exact atmospheric circulation may not be fully representative of reality in a minority of the simulations, the general steady-state thermodynamic solution appears to be robust. Regardless of the scale of the lake, the overall behavior of the atmosphere-lake system is similar, and this generally means that any size lake may be used to investigate the impact of other parameters on the system.  The 300 km-wide lake (sim #4) is used as the control scenario, because it provides a greater number of domain points over the lake and also exhibits the highest degree of symmetry.   Table 2 shows that many parameteric studies were conducted for a 100 km lake and a handful for the 32 km lake. It is not feasible to display the results of all these simulations, but the data for all the simulations are provided as supplementary material in NetCDF format.  Analyses of all these data demonstrate that the 300 km lake is indeed sufficiently representative of the results for the smaller lakes.

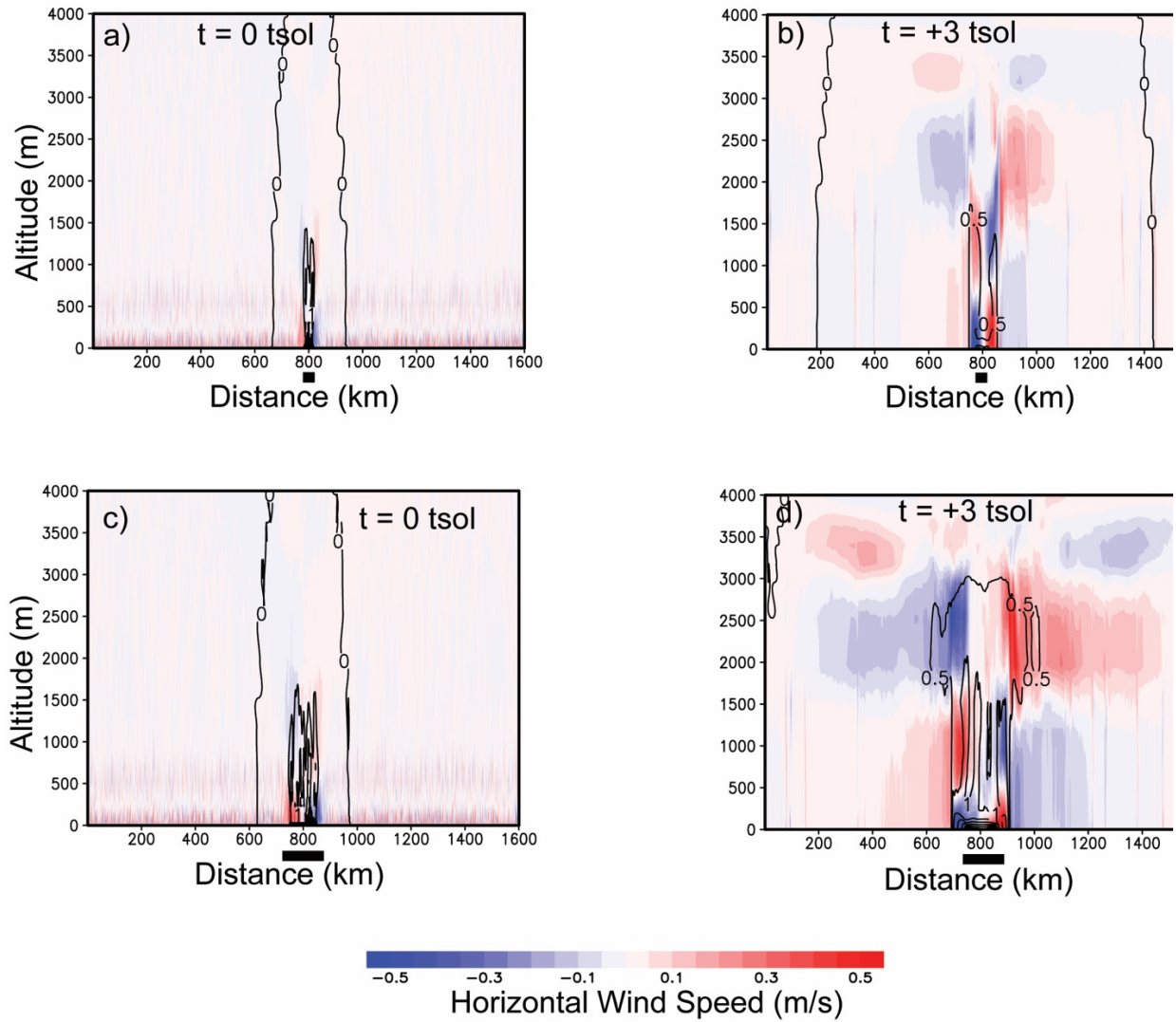

*Figure 12. Sea breeze evolution for a 32 km (sim #2, top row) and a 100 km (sim #3, bottom row) lake. Horizonal velocity is shaded with red (positive) values indicating flow from left to right and blue (negative) values indicating flow from right to left. Contours are methane mixing ratio (g/kg). The lake domain is shown by the black bar below the horizontal axis.*

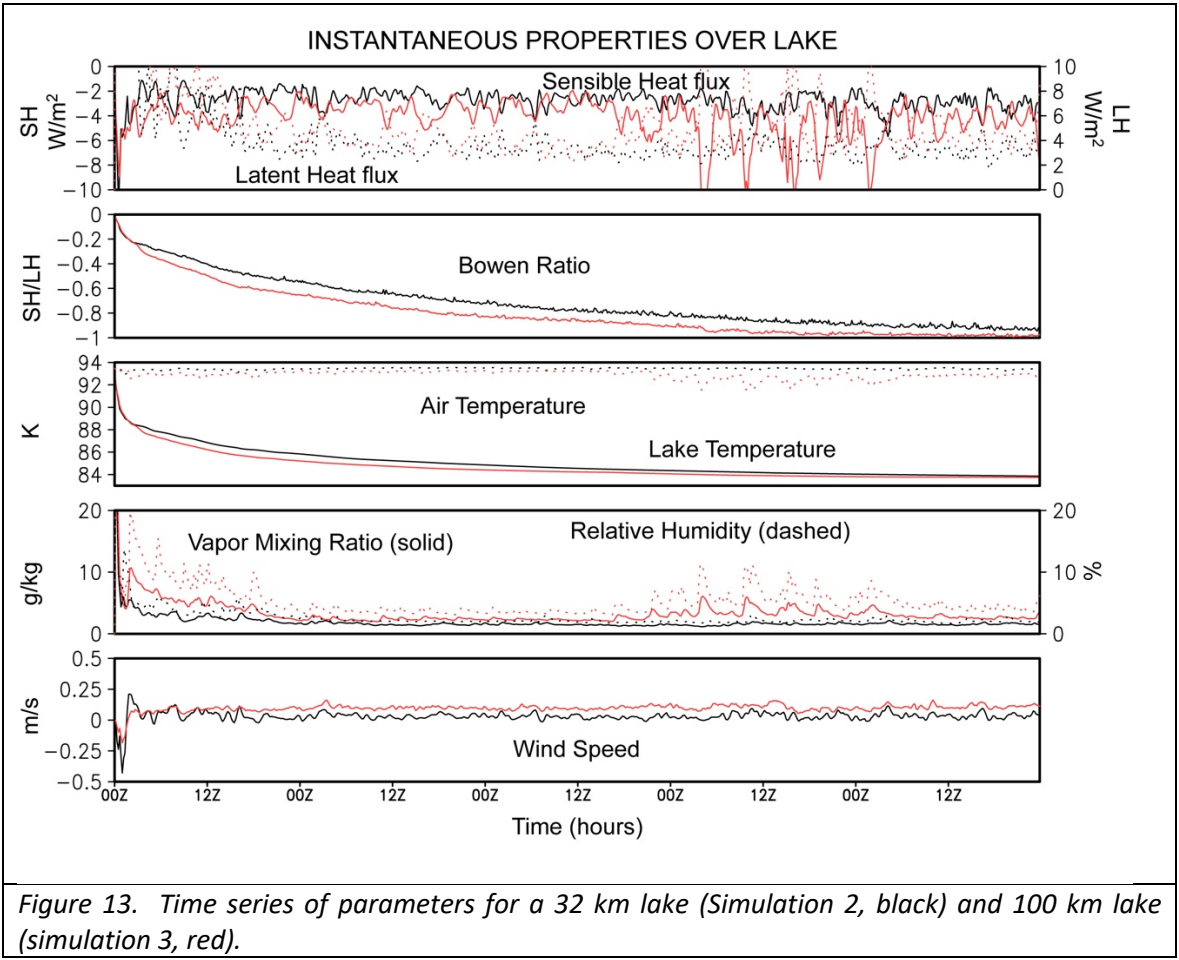

*Figure 13. Time series of parameters for a 32 km lake (Simulation 2, black) and 100 km lake (simulation 3, red).*

## 4.2. Effect of Lake Mixed Layer Depth

Per Fig. 3, the cooling of the lake depends on the assumed depth of the mixed layer, all other things being equal. Simulations 2, 3, and 4 exchange energy over a very shallow 1 m lake depth, and this drives a large response in lake temperature that in turn feeds back to the fluxes. A cold lake should produce a relatively large sensible heat flux due to the large sea-atmosphere temperature gradient while having a relatively small latent heat flux due to the temperature-dependent decrease in saturation vapor pressure. A modification of the evolution of the circulation might be expected as the lake mixed layer depth is changed. The initial plume circulation should remain mostly unaltered, because the large evaporation driving that plume will have a diminishing effect on lake cooling as the mixed layer depth increases. With less lake cooling and a corresponding lower sensible heat flux, the atmospheric cooling should also be reduced, and this should slow the development of the opposing sea breeze circulation.

Simulations 23, 24 and 25 test the effect of different mixed layer depths for a 300 km lake scenario (D=10 m, 100 m, and 30 m, respectively), and the time series of key physical variables is shown in Fig. 14 (compare to Fig. 10). The results are consistent with expectations. For the largest mixed layer depth (100 m, sim #24), the cooling of the lake is much slower and extends the time over which a dominant plume circulation is present. The latent heat flux is larger than the sensible heat flux over the duration of the simulation, and the Bowen ratio never reaches -1.0 as it does in the shallow mixed layer case. The sensible heat flux is closer to 5 $Wm^{-2}$ while the latent heat flux slow relaxes toward a similar magnitude. For all the mixed layer cases, the trend is toward a flux equilibrium condition, but extrapolation of the Bowen ratio trend indicates it would take a very large number of sols, perhaps dozens if not hundreds of sols, for the deepest mixed layer case. As a result, the lake temperature continues to cool in all the cases, and the cooling time constant becomes longer as the mixed layer depth increases. Once the initial plume circulation starts to dampen and the sea breeze ramps up, winds drop to fractions of a meter per second. The average air temperature drops ~1 K, but there is significant time variability. The vapor mixing ratio in the marine layer is two to three times higher than in the canonical case. The saturation mixing ratio at the temperature of the air is close to 70 g/kg, so, despite the relatively large values, the air is still well below saturation. The deepest mixed layer case (100 m, sim #24) has the greatest moisture content, which might be expected since the lake is the warmest and the latent heat flux is higher, especially during the early plume circulation phase.

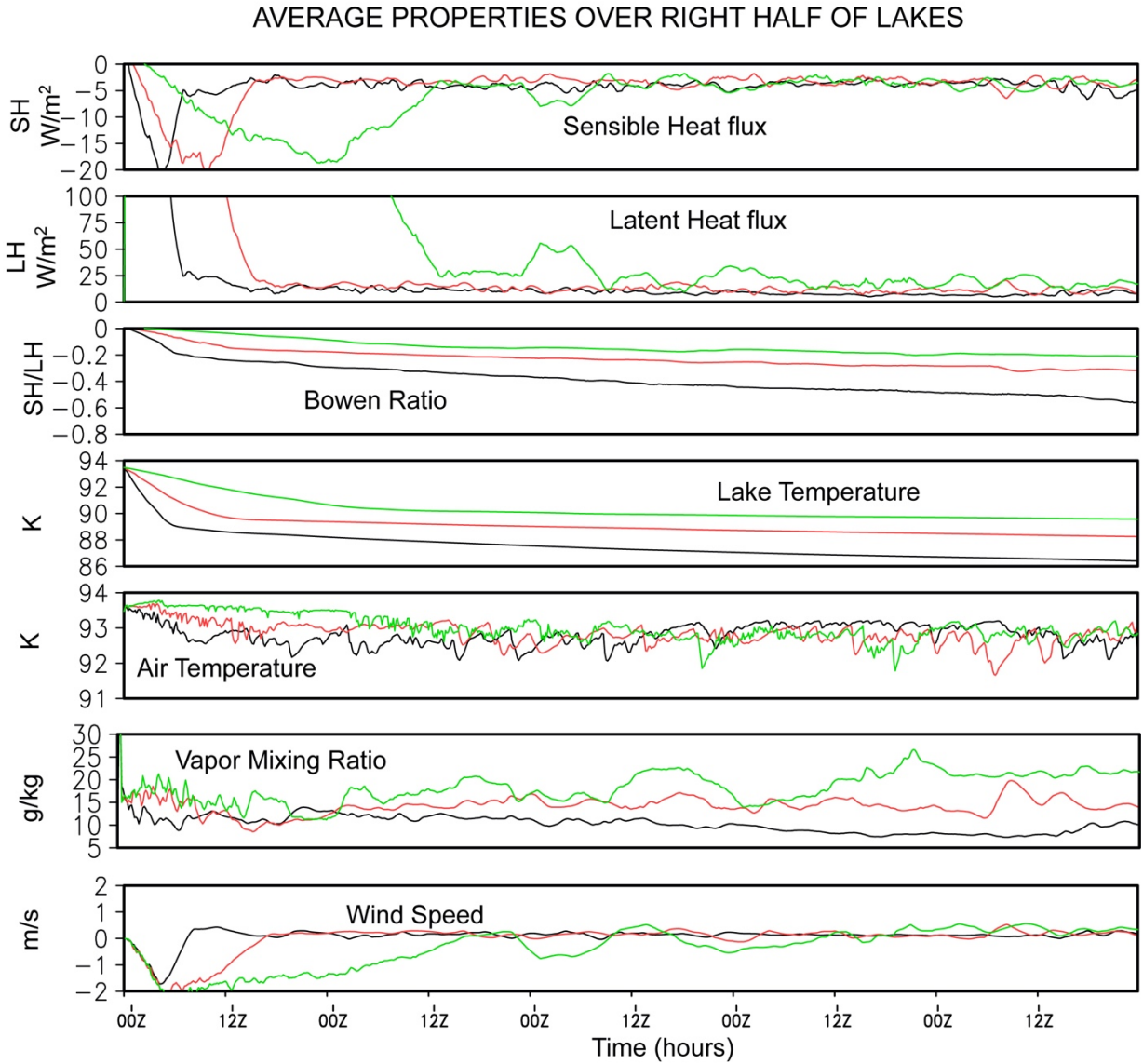

*Figure 14. Time series of key physical variables for a 10 m (Sim #23, black), 30 m (Sim #25, red), and 100 m (Sim #24, green) lake mixed layer depth.*

The wind and vapor field for the 10 m mixed layer depth case (sim #23) at 5 tsols, averaged over 6 hours, is shown in Fig. 15. The robust and dominant plume circulation is obvious, as is the marine layer and sea breeze. The transition zone is clearly evident and is substantially wider than in the canonical case. Although there is a strong change in properties at the sea breeze front, it is clear that the influence of the lake has propagated ~300 km inland. A close up view of the low-level circulation within the transition zone shows a series of weakly alternating velocity structures between the lake shore and the plume circulation inflow (Fig. 15 inset).

There are some other key differences between the deep mixed layer case (100 m, sim #24) and shallow canonical case; the same basic structures are all present, but the details do differ. The divergent outflow from the sea breeze extends to over 3.5 km and actually exceeds the depth of the plume circulation which reaches to ~2.5 km in altitude near the shore but is shallower over land (not shown, see supplementary data). This is curious, since the initial plume circulation is strongest and longest-lived in the deep mixed layer case. A more careful examination shows that the early initial plume circulation is, in fact, quite deep—up to 5 km, but it descends with time as the sea breeze develops. The gradually cooling air partially offsets the virtual buoyancy effect so that the overall plume buoyancy decreases with time. The marine layer is far moister in the deep mixed layer case, as was suggested by the time series, and generally speaking the air is slightly cooler than shallower cases (Fig. 14). Both the higher moisture and cooler air is indicative of a stronger marine layer even though the flux equilibrium condition has not yet been met. The moisture is also far greater aloft, by almost a factor of 10, which indicates that deep lakes are far more efficient at exporting vapor than the shallower lakes, which was expected from basic physics arguments.

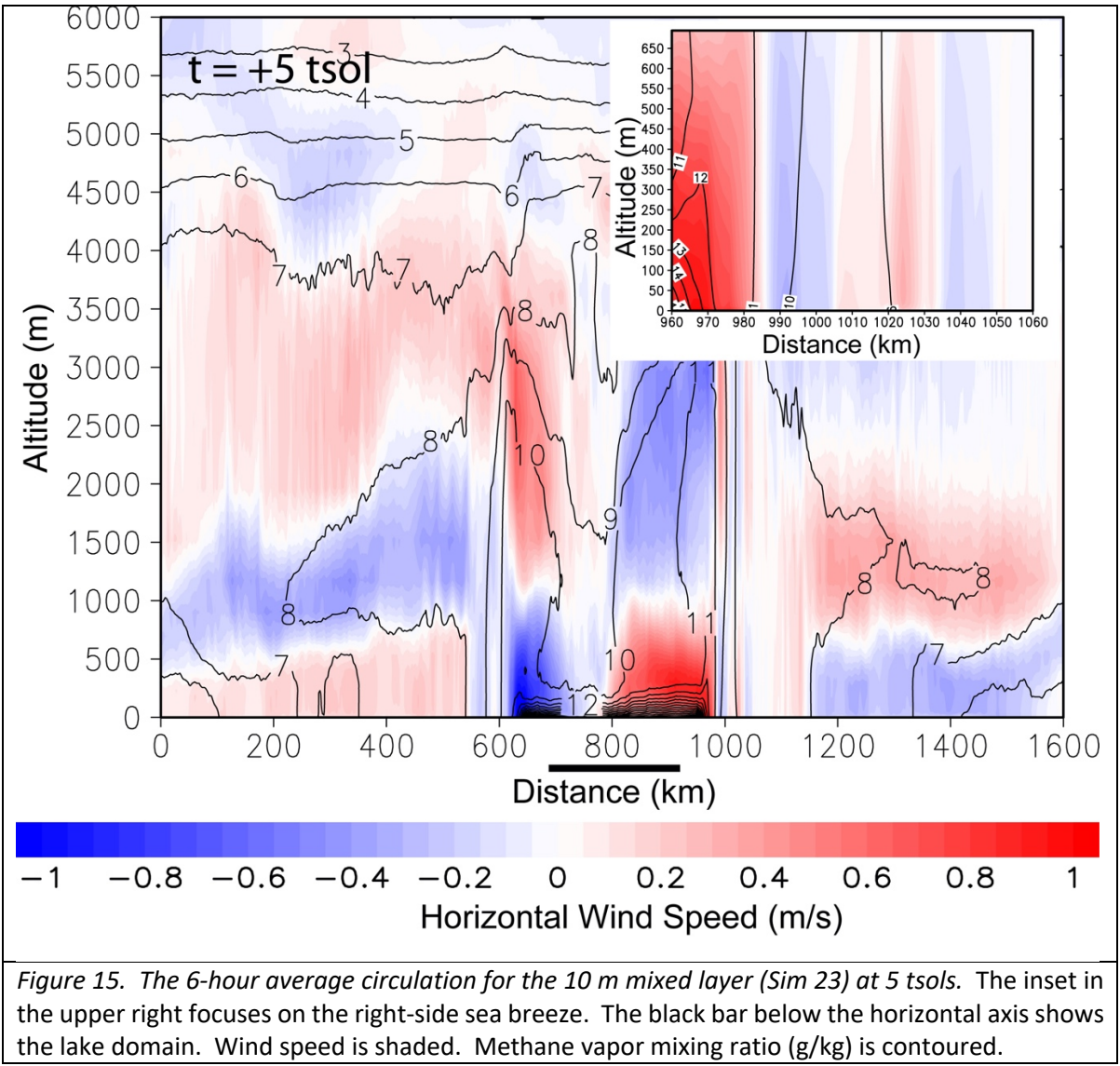

*Figure 15. The 6-hour average circulation for the 10 m mixed layer (Sim 23) at 5 tsols.* The inset in the upper right focuses on the right-side sea breeze. The black bar below the horizontal axis shows the lake domain. Wind speed is shaded. Methane vapor mixing ratio (g/kg) is contoured.

It is clear that the lake temperature has a strong influence on the equilibrium vapor content of the marine layer. A warm lake pumps more methane into the air than a cold lake. If actual Titan lakes and atmosphere are in thermal equilibrium but in large vapor disequilibrium, deeper mixed layer lakes may be far more effective at pumping methane into the atmosphere of a long equilibrium time constant than shallow mixed layer lakes.

An immediate question follows from these results: What is a typical or representative mixed layer depth? On Earth, water has a maximum density at ~4 °C. Cooling of warm water will result in negatively buoyant parcels sinking until they reach their thermal equilibrium level, entraining warm water along the way. Thus, the depth of an ocean or lake mixed layer depends on the thermal, or equivalently, the density profile as a function of depth. If, for example, a water lake has an isothermal profile well above 4 °C, a cooled surface water parcel will sink to the bottom if there is no entrainment.

Pure liquid methane (or ethane) has a similar behavior to warm water. If a large Titan lake like Kraken Mare with a maximum depth of >100 m is assumed to be pure methane, a cooled surface parcel would be expected to sink to the bottom, or at least to a lesser depth determined by the amount of warm liquid methane entrainment. Over time, continually entraining fluid will ultimately mix to the bottom of the lake. Looking back to the 1 m mixed layer depth simulation (sim #4), the cooling of the mixed layer is significant (many Kelvins), and if a lake were deeper than the assumed mixed layer, it would be highly likely that the mixed layer depth would grow from the initial 1 m. Thus, the initial shallow mixed layer might be expected to evolve towards the 10 m scenario (sim #23), and if the process continued, to 30 m (sim #25), and 100 m (sim #24). In other words, assuming a reasonably deep lake, the evolution of the lake temperature in the 1 m mixed layer case would be unrealistic, because as the surface cooled that fluid would sink and would be replaced by warmer fluids. In turn, the convective overturn would reduce the sensible heat flux while enhancing the evaporatively-driven plume. Although a shallow mixed layer is most conducive to generating a sea breeze, it is simultaneously a self-limiting if not self-defeating process. A cold lake surface and shallow mixed layer are, for the most part, physically inconsistent. This is not all that different from Earth. Stable stratification of a lake is warm water over cold water (a thermocline) and not the other way around. Consequently, the 1 m mixed layer depth scenario should be very close to upper limit of the ability of the Titan air-sea system to generate a marine layer and sea breeze circulation over a pure methane lake.

The above argument holds for pure methane, but Titan's lakes are likely to have other solutes, especially ethane [Lorenz, 2014]. Depending on composition, it is possible that the lake liquids will exhibit a non-standard behavior similar to water below 4°C. Namely, the fluid may decrease in density as it cools [Tan et al., 2015]. In this case, a shallow mixed layer and cooling lake is physically consistent. The coldest fluids will remain near the top of the lake and even freeze if sufficiently cooled. In this case, the shallow mixed layer depth case of 1 m (sim #4), may be the most realistic. The actual behavior of an actual lake parcel on Titan is still an open question. Lake freezing is discussed in more detail in Section 6. Also, not all lakes are deep and even deep lakes can have shallows for which a shallow mixed layer is locally relevant.

## 4.3. Air-Lake Temperature Differential and Initial Relative Humidity

The simulations presented up to this point all suggest that the air-sea system will tend to evolve to a state where the lake is cooler than the air and to an atmosphere that is moist but not saturated. The time constant of this evolution depends strongly on the mixed layer depth, but it would be reasonable to expect that more realistic initial atmospheric background conditions are in thermal and vapor disequilibrium, rather than equilibrium, with the lake. In principle, all the simulations could be integrated over a very long time, but in many cases that would be hundreds of tsols, and the simulations would likely suffer from spurious boundary condition effects and other numerical instabilities. The solution is to initiate simulations with a lake colder than the atmosphere and with a moister atmosphere, both of which are possibly closer to reality. The intuitive expectation for this scenario might be for a diminished initial buoyancy plume circulation and an enhanced sea breeze circulation. Simulations 31-33 test the expectations for 300 km wide and 100 m mixed layer lakes initialized with different cold temperature deficits (sim #31 at 6 K, sim #32 at 4 K, and sim #33 at 6 K). Simulations 34-36 are 300 km wide lakes with a 1 m mixed layer and an "unstable" initial relative humidity of 20%, 50% and 80%, respectively. Finally, simulations 60-62 explore the solutions with "stable" initializations of a 20%, 50% and 80% relative humidity, but are otherwise identical to simulations 34-36. We note that the coldest lake cases are likely below the freezing point, but the point here is to test the response to increasingly large differences between the atmosphere and lake, and the cold lake cases serve that purpose; the very cold lake ($\Delta T = -6\ K$) represents a bounding case. The variable humidity experiments also test by proxy the effect of an impure methane lake. The addition of ethane or other volatile compounds decreases the saturation vapor pressure of methane via the solute effect. Moistening the atmosphere decreases the vapor pressure gradient between the atmosphere and the lake, which is essentially the equivalent of adding contaminants to the lake. Eq. (1) is agnostic to the origin of the vapor difference between air and sea, so moistening the atmosphere or adding contaminants to the lake produce the same effect

In the two coldest 100 m mixed layer cases (sim #31 with an initial 6 K deficit and #32 with an initial 4K deficit), a robust sea breeze develops quickly with very little evidence of a plume circulation. The representative solution is shown for Simulation 31 in Fig. 16. The sea breezes slowly propagate inland. In contrast, the lake initialized with a 2 K deficit (sim #33) shows a plume circulation and weak evidence of a sea breeze. Thus, for the deep mixed layer case, there is a transition from a plume-dominated circulation to a sea breeze-dominated circulation as the lake cools.

The results of the cold lake experiments are roughly in accordance with intuitive expectations. With an initial cold lake, the initial latent heat fluxes are smaller than in most previous cases, because the saturation vapor pressure over the colder lake is lower. The initial sensible heat flux tends to be larger, because the simulation starts off with an imposed temperature gradient between the air and lake. There is a moistening of the atmosphere due to evaporation, but it is much smaller than when the lake and atmosphere are initially at the same temperature. If the lake is cold enough, the cooling of the air through the sensible heat flux is sufficient to both initiate a sea breeze and reduce the virtual buoyancy effect such that the plume circulation is suppressed (Fig. 16a). The simulation then proceeds toward a fully mature sea breeze circulation with little to no evidence of a plume circulation at later times (Fig. 16b). In the slightly cold lake case (2 K deficit, sim #33), the virtual buoyancy effect dominates, and a plume circulation is initiated (Fig. 16c). A sea breeze does not initiate until the lake cools later in the simulation (Fig. 16d). The rate of the lake cooling is relatively slow, because of the depth of the mixed layer, thus the plume circulation can persist for some time, and remain as a background circulation once the sea breeze does develop.

We note that in nearly all the 100 m mixed layer cases with an initially cold lake, the lake temperature tendency associated with the net turbulent flux is so small that it falls within machine error. Additional code was added to accumulate these tendencies until they were large enough to be used to update the lake temperature on a longer timestep, as previously discussed. Without this code addition, the lake wouldn't change temperature even though the fluxes are non-zero. Once again, the implications of the weak forcing on numerics is discussed in Section 9.

The time series of relevant physical parameters for the cold, deep mixed layer cases (Fig. 17) provide greater insight. The sensible heat flux for the warmest of the three lakes (sim #33 with a 2 K initial deficit) ramps up quickly even though it starts with the smallest air-lake temperature gradient. This is because Simulation 33 rapidly develops a plume circulation so that wind speeds increase to nearly 1.5 m/s. Also, the stability of the lowest layers is less than in the cold lake cases, which will result in a greater bulk transfer coefficient. The latent heat flux in Simulation 33 also increases dramatically for the same reasons, and the lake cooling trend is initially greatest in this case. The two colder lakes (sim #31 with an initial 6 K deficit and sim #32 with an initial 4 K deficit) show a much more muted response. The colder lakes suppress the latent heat flux. The small flux of vapor is insufficient to produce a strong virtual buoyancy circulation, and no plume circulation develops. The small latent heat flux also means that the lake cooling is very slow. Due to the lack of a plume circulation, the winds remain small.

None of the simulations achieve a flux equilibrium condition, although the monotonic trend towards that solution is clear. Nevertheless, when the air and lake are out of thermal equilibrium, the time over which a flux equilibrium condition may be achieved increases as the depth of the mixed layer increases. This finding is not entirely unexpected, but it is possible to envision a scenario in which the equilibrium time constant becomes so long that equilibrium is never achieved, for all practical purposes. Lake cooling is also very slow for the deep mixed layers, and given the small fluxes in the later part of the simulation, the lake temperatures are nearly constant with time. Interestingly, the air temperatures are very similar in all the cases regardless of the underlying lake temperature. Thus, the coldest lake tends to have a slightly higher magnitude sensible heat flux. The near-surface wind speeds in the latter half of the simulation are very small (nearly zero), although the winds aloft are still present (Fig. 16).

With smaller, 100 km wide lakes (sim #28, 6 K deficit; sim #29, 4 K deficit; and sim #30, 2 K deficit) that are otherwise similar to their wider lake counterparts, the solutions are in many ways almost indistinguishable (not shown). Lake size does not play a major role in the overall solution in these scenarios.

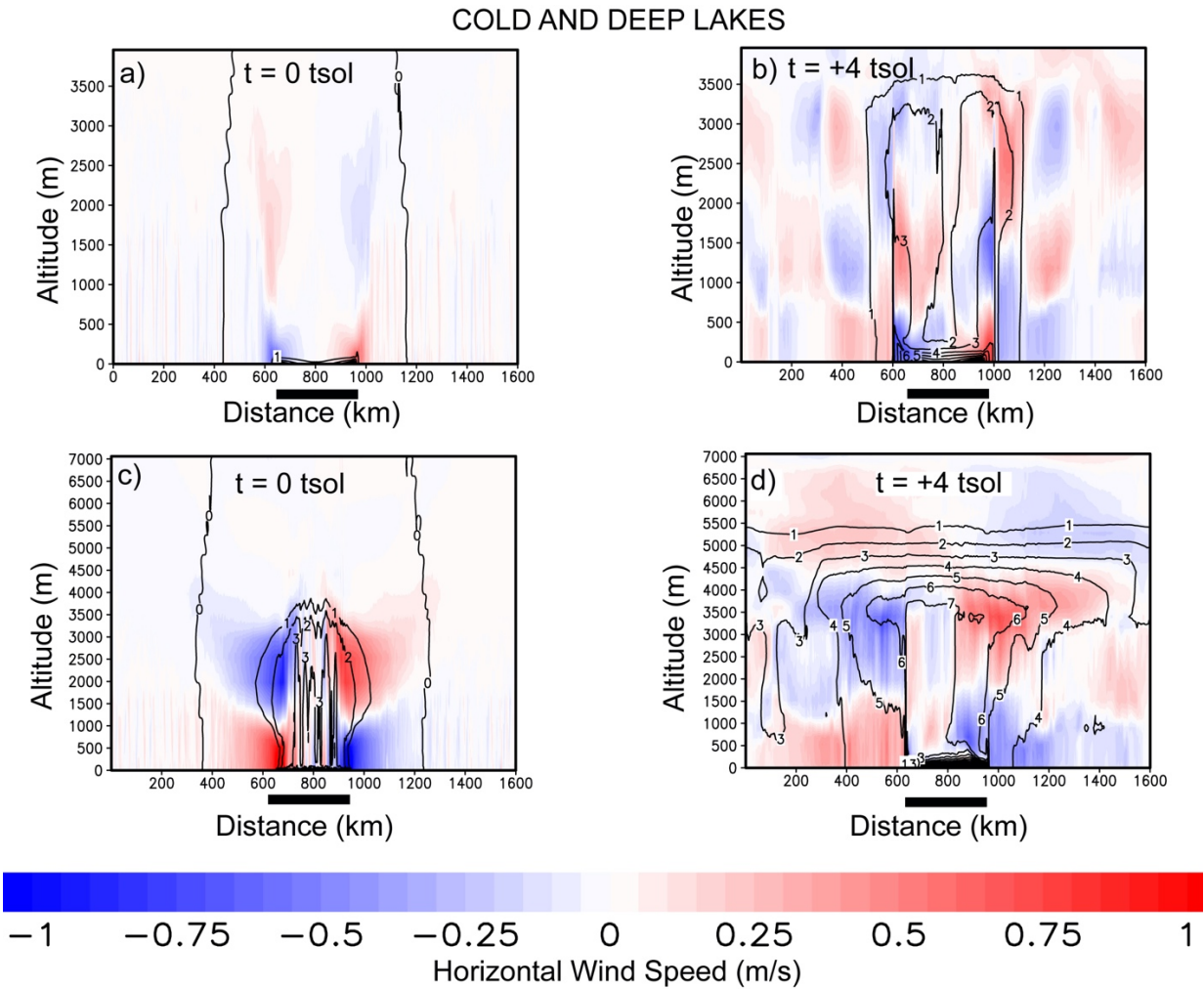

*Figure 16. Results from deep mixed layer cases with a cold lake. Sim 31 (coldest initial lake with a 6 K deficit, top) and sim 33 (initial 2 K deficit, bottom). The initial circulation depends strongly on the lake temperature deficit with a sea breeze for the very coldest case and a plume circulation for the cold lake. The solution 4 tsols later reflects the initial circulations with the strongest background plume circulation found in the cold scenario and almost no plume found in the very cold case. Note the different vertical scales between the two cases. The black bar below the horizontal axis shows the lake domain. Wind speed is shaded. Methane vapor mixing ratio (g/kg) is contoured.*

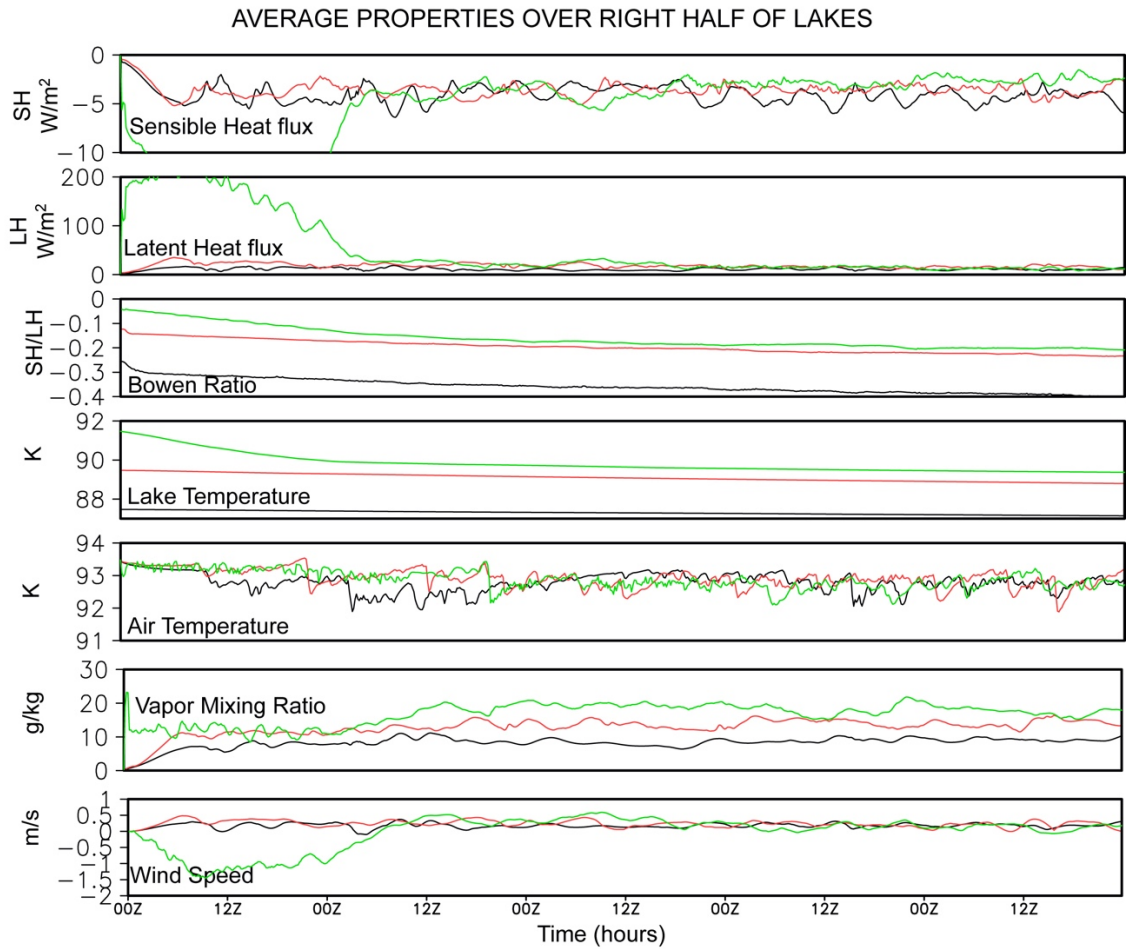

*Figure 17. Time series of key physical variables for the cold and deep lake simulations 31 (black), 32 (red), and 33 (green).*

Experiments 54, 55, and 59 are identical to 31, 32, and 33, but with a shallow mixed layer (1 m vs. 100 m). The shallower mixed layer results in a larger lake temperature tendency for a given latent heat flux compared to the 100 m cases, which eliminates the machine precision issue. The two coldest 1 m mixed layer cases (sim #54 with 6 K and sim #55 with 4 K) are dominated by a sea breeze, as was the case for the deep mixed layer (Fig. 18). The cold lake wit a 2 K deficit (sim #59), however, deviates from its companion deep mixed layer result (sim #33). Instead of a dominant plume circulation, a sea breeze circulation is well established and superimposed on a relic plume circulation.  This is explained by the much more rapid cooling of the lake in Simulation 59 as a result of the shallow mixed layer.  Thus, when the mixed layer is shallow and cold it is far more difficult for a dominant and strong plume solution to develop; a sea breeze develops more quickly than seen in the deep and cold scenarios.

In all the shallow, cold lake cases, the average wind over the lake trends very close to calm.  Even so, the fluxes are nonzero by the end of 5 tsols and appear to reach a quasi-steady state. The lake temperatures are also nearly stabilized.  While lake cooling is taking place, the fluxes are small enough that the rate of cooling is almost negligible.   From an air-sea exchange point of view, the end results of all the shallow cold cases are nearly indistinguishable except for the lake with the smallest, 2 K initial air temperature differential (sim #59).

## COLD AND SHALLOW LAKES

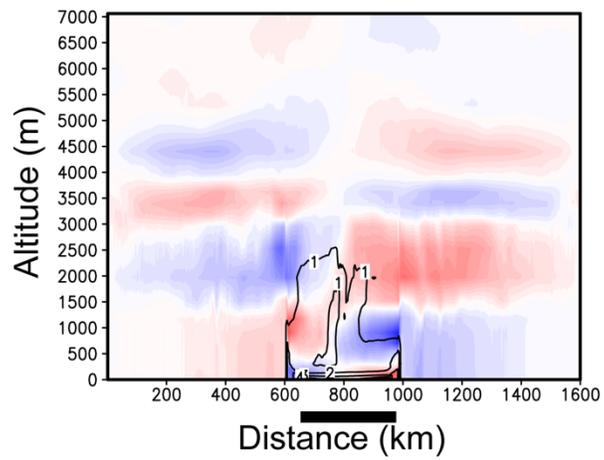

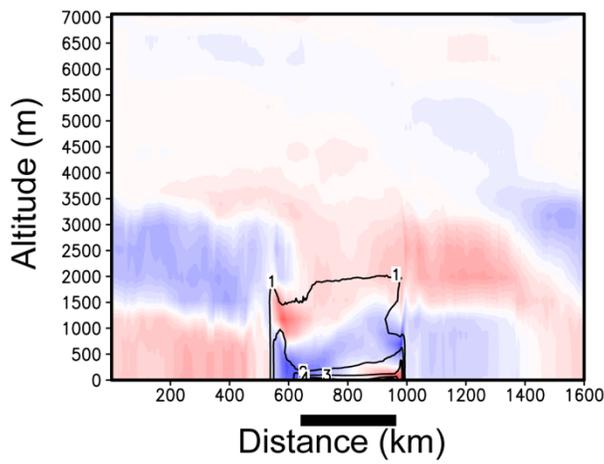

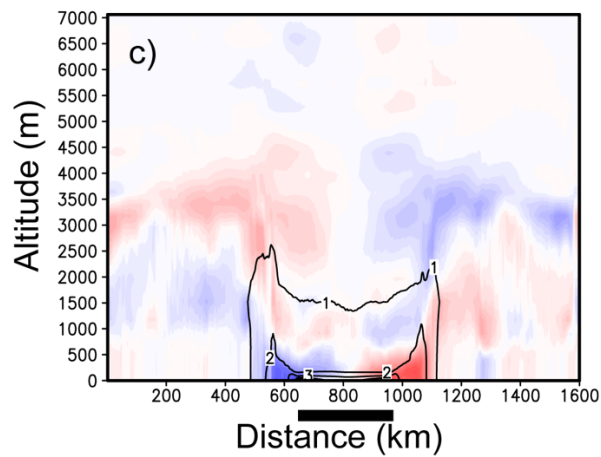

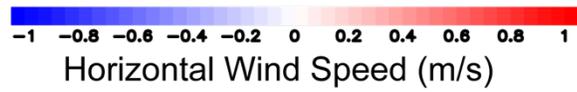

*Figure 18. Shallow, cold lake scenarios at tsol 5. Sim 54, initalized 6 K colder than the air (a); Sim 55, 4 K colder (b); Sim 59, 2 K colder (c). The sea breeze becomes more dominant and the plume circulation less dominant as the initial lake is initialized colder. The black bar below the horizontal axis shows the lake domain. Wind speed is shaded. Methane vapor mixing ratio (g/kg) is contoured.*

The effect of increasing atmospheric humidity (using the "stable" initial conditions of Simulations 60, 61, and 62) is in concert with the intuitive expectations (Fig. 19). As initial atmospheric humidity is increased, the initial evaporation rate is decreased and the resultant plume circulation is diminished. In fact, there is little to no evidence of the plume circulation at early times and only a weak background circulation consistent with a plume circulation is visible later in the simulation. Another factor minimizing the plume circulation is that the virtual buoyancy is smaller compared to an initial dry atmosphere. The reduced evaporation rate also means the lake cools more slowly, the magnitude of the stable marine boundary layer is reduced, and the strength of the sea breeze circulation is reduced. The asymmetry above the shallow sea breeze at later times in the simulations is noticeable, especially in the highest humidity case. The highest humidity scenario tends to have the weakest net forcing. The unstable initial conditions (Simulations 34, 35, and 36) are not presented here, but show the same general result. The unstable results are available in the supplementary material.

All the stable humid simulations (simulations 60-62) reach the flux equilibrium condition in less than 3 tsols. The time series of parameters for the humid cases are not shown, but the data are available in the supplementary material. The magnitude of the fluxes at equilibrium are < 2 W m$^{-2}$, and the smallest fluxes are found in the most humid scenario. The Bowen ratio is ~1.0 after a little over 2 tsols. Winds are also very small (< 10 cm/s), but the most humid case shows some of the greatest variation (± 2 cm/s). This may very well be due to asymmetries and nonphysical computational modes shining through the very weak forcing and circulation. The lake temperatures equilibrate quickly and become effectively constant with time. The most humid case (sim #62 with 80% RH) has very little lake cooling (<1 K), while the least humid case (sim #60 with 20% RH) has a lake that cools to ~86 K. The atmospheric methane content is nearly unchanged from the initial condition. The humid simulations come close to an equilibrium solution that is also the trivial solution of fluxes and wind nearly equal to zero. Importantly, however, there is still an exchange, and winds are non-zero aloft even if they are very small near the surface.

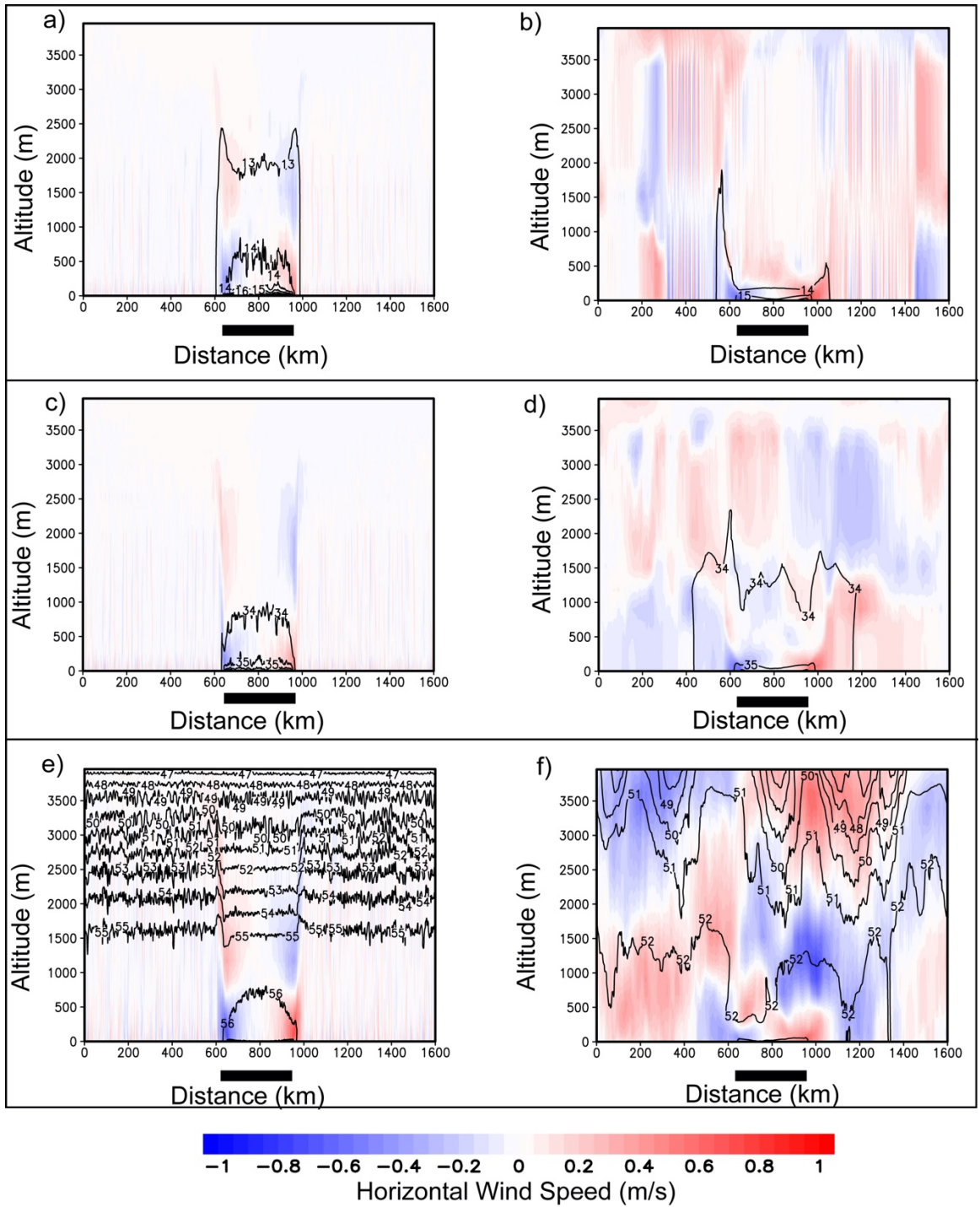

*Figure 19. Initial and +4 tsol circulations for the 20% (top), 50% (middle) and 80% (bottom) RH cases (Sims #60-62). Note the asymmetry at 4 tsols. The black bar below the horizontal axis shows the lake domain. Wind speed is shaded. Methane vapor mixing ratio (g/kg) is contoured.*

As previously described, an increase in atmospheric humidity can be viewed as a proxy for a reduction of saturation vapor pressure over the lake associated with a mixture of solutes. There is a reasonable expectation that lakes are not pure methane and instead have some non-negligible amount of liquid ethane and perhaps other volatile organics [Lorenz, 2014]. As the fraction of liquid methane is reduced, the methane saturation vapor pressure over the lake will be reduced, and thus the intensity of the initial plume circulation will be reduced due to a reduction in the lake to air vapor gradient. The reduction in the vapor gradient is analogous to increasing the initial atmospheric humidity. Ethane and higher order organics have a higher molecular weight than methane and nitrogen, so the virtual buoyancy contribution of these volatiles, which is not included in these simulations, counters that of methane. The concentration of these heavier gases, however, would almost certainly be insufficient to substantially reduce the methane buoyancy. On the other hand, it is possible to imagine a nearly pure ethane lake with evaporative cooling and negative virtual buoyancy. Under such conditions, only an ethane vapor-rich sea breeze circulation would be possible.

## 4.4. Effect of Background Wind

With the exception of the convective storm environment [Barth and Rafkin, 2007; Charney et al., 2015; Rafkin and Barth, 2015], Titan is thought to have a sluggish atmosphere with large-scale surface winds of ~1 m/s or less [Bird et al., 2005; Tokano et al., 2009]. The mtWRF-simulated near-surface winds in and around the lakes are also generally small (~1 m/s or less), but it is not unreasonable to expect that the air-sea interaction circulations could be embedded in a large-scale mean wind environment that is at least of comparable magnitude to the local circulations. On Earth, mean winds distort sea breeze circulations [Leopold, 1949; Blanchard and López, 1985; Finkele, 1988; Atkins and Wakimoto, 1997; Miller et al., 2003; Gilliam et al., 2004; Crossman and Horel, 2010]. Sea breeze fronts propagating in the direction of the mean wind tend to push faster and more deeply inland.  In contrast, a sea breeze opposed to the mean wind moves inland more slowly and can even become locked to the coastline or remain entirely offshore.

Many simulations were conducted to explore mean background winds of 1 m/s and 3 m/s with either periodic or open boundary conditions, with different mixed layer depths, and with different initial lake temperatures (e.g., sims #8, 9, 26, 27, 56, and 57).  Although small by Earth standards, 1 m/s and 3 m/s are probably a stiff breeze for Titan and are similar in magnitude to the previously discussed local lake circulations when taking into account the frictional slowing of the mean wind near the surface. Even though the initial background wind is specified as a constant, friction reduces the wind speed near the surface with the greatest decrease over the land where the roughness is higher.  Thus, the imposed mean wind condition is really a constant wind "above the deck" with a frictional decrease in wind speed toward the surface.

Periodic boundary conditions (e.g., sims #26 and #27) are somewhat problematic in that they allow the circulation to blow downwind and then re-enter the domain on the other side.  At 1 m/s, an air parcel will move through the domain in about 1 tsol.  Once the simulation time exceeds the Lagrangian advective time scale, the simulations are representative of an infinite chain of lakes rather than an isolated lake.  At the same time, it is reasonable to argue that the northern lake district could be roughly characterized as a series of lakes that are potentially downwind of one another. In this case, periodic boundary conditions allow for the simulation of multiple circulations interacting with one another.

Open boundary conditions allow circulations to exit the domain, but at the possible expense of spurious boundary condition effects on the interior solution. Indeed, this was seen in Simulation 57 after about 2.5 tsols (Fig 20; also see discussion in §9). Spurious circulations from the boundary push into the domain as the real circulations exit the domain. These numerical, boundary condition artifacts produce asymmetries that can overwhelm the physical solution. The effect often seems to be triggered by the arrival of the plume outflow circulation at the boundary, and the solution degrades quickly thereafter. Prior to 2.5 tsols, the solution appears largely unaffected. Even without spurious boundary noise, once a system exits the domain, there is no longer any numerical information on the structure or evolution of that system. The exiting of the system (e.g., a sea breeze front) can be mitigated by increasing the size of the domain, but that solution quickly becomes computationally impractical. Simulations with an imposed constant backgound wind suffer most from this problem, since the mean wind tends to advect circulations in a preferred direction.

The deep mixed layer simulations with an imposed background wind (e.g., sims #56 and #57) have very slowly cooling lakes, as expected based on previous simulations. The latent heat flux is large enough that the lake temperature tendencies are generally well above machine precision, unlike the zero-wind, cold lake cases (sims 31-33). The result from Simulation 56 (1 m/s with open boundary conditions) is shown in Fig. 20, as a representative example of the deep mixed layer results. Simulation 57 (3 m/s with open boundary conditions) is also shown, but recall that the tsol=4 result appears to show a strong non-physical solution. The initial plume circulation is difficult to see because it is not sufficient to counter the imposed mean wind, but wind perturbations are consistent with a weak plume circulation superimposed on the stronger background wind. The circulation is tilted slightly downwind. A moist, but not very cold marine layer forms with the plume circulation.

By four tsols, the plume circulation is more obvious. In Fig. 20b, the left side of the plume circulation extends from the left boundary to approximately x=400. Within this region, the inflow is faster than the initial mean wind as a result of constructive interference. The outflow in this region is directed opposite to the imposed wind, and is nearly as strong in magnitude as the inflow. In this case, it appears that the outflow strength is required by mass conservation to roughly balance the inflow and effectively counters the opposing mean wind. On the right side of the domain, the plume circulation has almost exited the domain. In the deep mixed layer case shown in Fig 20d, the stronger mean wind has already pushed the plume circulation out of the right side of the domain by 4 tsols, and the resulting boundary noise has fully propagated into and contaminated the solution.

Simulations 65 (1 m/s) and 66 (3 m/s) have shallow mixed layers (1 m), and were designed to test the effect of the mean wind in cases where a mature, cold marine layer with a well-defined sea breeze should form more quickly than in the deep mixed layer cases. In Simulation 65 with the weaker wind, the sea breeze spins up quickly due to the rapidly cooling lake. Within the first two hours (Fig 20c), a slowing of the winds near the sea breeze front of the marine layer is apparent. By later times (Fig 20d), the sea breeze is fully established, and a background plume circulation is also present. The winds associated with both the sea breeze circulation and the plume circulation show an asymmetry that slightly favors the mean wind, and the final result looks very similar to the result in the deep mixed layer case (Fig. 20b) The primary difference between the deep and shallow cases with the 1 m/s wind (sim #56 vs. sim #65) is that the sea breeze circulation is stronger on the right side in the shallow mixed layer scenario (sim #65). This is not unexpected since the shallow mixed layer allows for the faster spin up of the sea breeze.

The 3 m/s mean wind case with a shallow mixed layer (Sim #66) remains stable unlike the deep mixed layer case (sim #57). No doubt, it is because of the quick establishment of a marine layer and the suppression of the plume circulation. Without a well-defined plume circulation to advect out of the domain, there is no forcing of non-physical solutions at the domain boundary. The initial circulation is tilted downwind (Fig. 20g) similar to the deep mixed layer case (Fig. 20e). By four tsols (Fig 20f), there is little evidence of either a plume or sea breeze circulation. However, if perturbation winds are examined (not shown), there is a hint of a sea breeze circulation. The moist marine layer is seen to extend substantially downwind near the surface, although the moisture content decreases with distance from the lake. There is effectively no marine air penetrating upwind. It appears that a strong enough wind effectively overwhelms both plume and sea breeze forcing. Winds near the surface, even upwind, are weaker than the imposed background wind, but that is due to friction.

Simulation 67 was designed to test the effect of an initial cold lake (2 K deficit) under an imposed wind background (3 m/s). With an initial cold lake, the initial plume circulation is strongly suppressed, while the cold marine layer and sea breeze develop even more quickly than in the shallow case alone (sim #66, not shown). As a result, the solution at 4 tsols differs from just the shallow case in that a very week sea breeze can develop, but it is otherwise not strongly different in character.

The time series of physical parameters for Simulations 56 (1 m/s with 100 m mixed layer), 65 (1 ms/ with 1 m mixed layer), and 66 (3 m/s wind with 1 m mixed layer) are shown in Fig. 21. Simulation 57 (3 m/s with deep mixed layer) is not shown, since it is contaminated with boundary noise after only a couple of tsols. Despite the difference in wind speed between 65 and 66, the results of these two simulations look very similar. Both have a latent heat flux that rapidly drops, driving the Bowen ratio to a value of near - 1.0 in less than two tsols. The lake temperature stabilizes to a cold ~84 K as a result. Unlike prior simulations with no mean wind, the magnitude of both the latent and sensible heat flux is larger—10 Wm$^{-2}$ or greater. The air temperature also tends to be colder than in prior simulations. Over the lake, the wind remains less than 1 m/s, regardless of the background mean wind, but it is also generally greater than in all the no wind scenarios. The air temperature also drops over time more than is generally seen in the no wind cases.

Simulation 56 (1 m/s with 100 m mixed layer) exhibits many of the characteristics previously seen in lakes with deep mixed layers. The latent heat flux is very large for an extended period of time. This result is due to the slow cooling of the deep lake which keeps the saturation vapor pressure over the lake relatively high and does not allow for the fast generation of a cold, stable marine layer. The sensible heat flux is also in line with prior deep mixed layer cases. The small temperature differential between the lake and atmosphere limits the sensible heat exchange. The Bowen ratio magnitude remains small and is only ~-0.2 after 4.5 tsols. A value of -1.0 would take 100 or more tsols to achieve based on a linear extrapolation of the trend. Even though the sensible and latent heat fluxes remain out of balance, the overall value of the sensible heat flux does drop by orders of magnitude toward the end of the simulation. The vapor abundance is higher compared to the two other mean wind cases. This is due to the warmer lake as well as the weaker marine layer that would otherwise tend to reduce the fluxes. Perhaps the most interesting result is that the circulations in simulation 56 and 65 look similar (Fig. 20) even though the time series of relevant parameters (Fig. 21) do not.

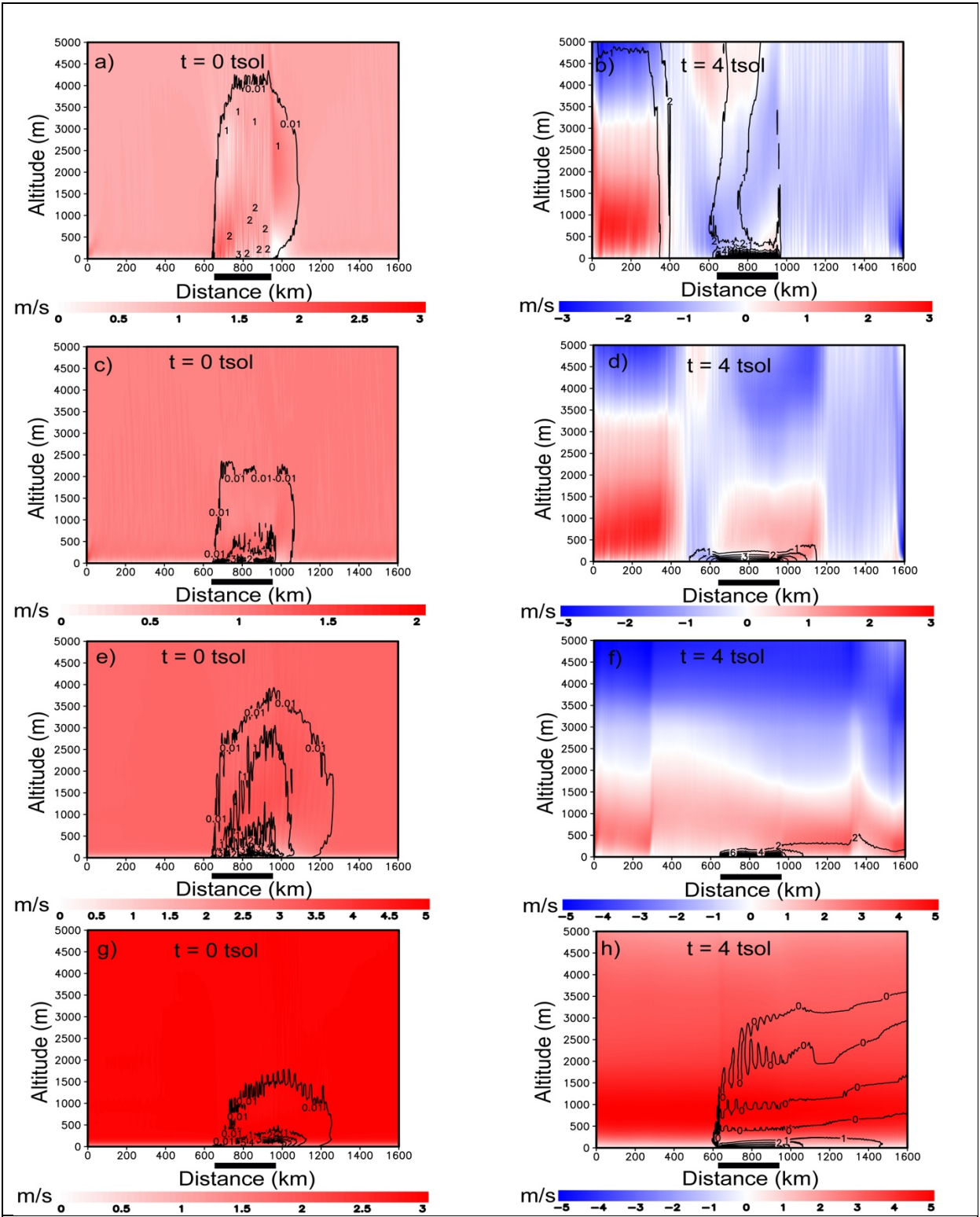

Figure 20. Results from mean wind simulations 56 (1 m/s, 100 m mixed layer, top row), 65 (1 m/s, 1 m mixed layer, second row), 57(3 m/s, 100 m mixed layer, third row), 66 (3 m/s, 100 m mixed layer, last row). Data are averaged over a two-hour window to remove transients. The black bar below the horizontal axis shows the lake domain. Wind speed is shaded. Methane vapor mixing ratio (g/kg) is contoured.

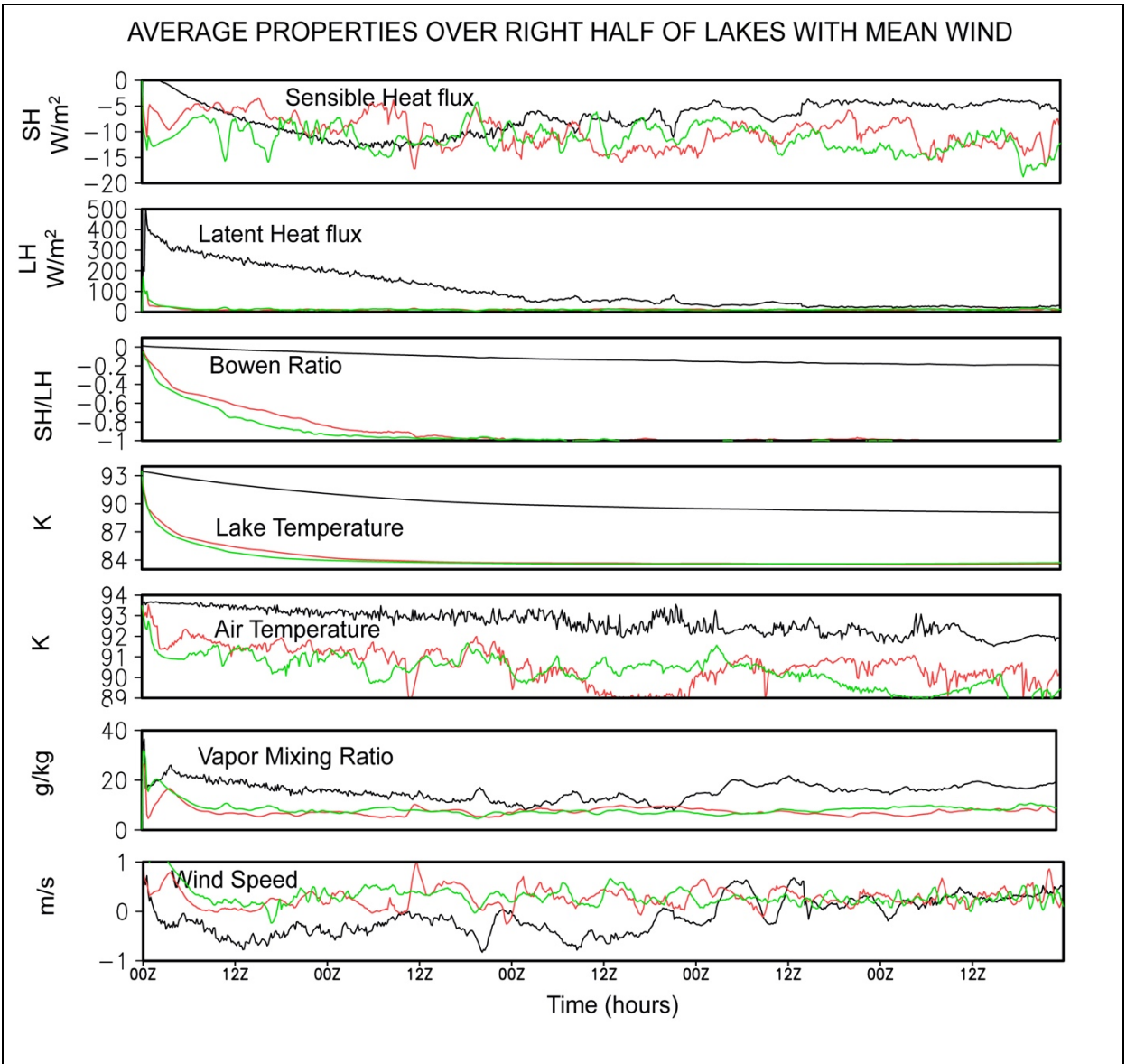

*Figure 21. Time series of relevant physical parameters for Simulations 56 (black), 65 (red), and 66 (green). With a shallow (1 m) mixed layer and wind (Simulation 65 red, and 66 green), the lake cools very quickly and the air temperature follows. Even though the latent heat flux in Simulation 56 is large for several sols, the depth of the mixed layer (100 m) tempers the rate at which the lake cools. Simulation 56 does not achieve a Bowen ratio anywhere close to 1.0 even though the lake temperature is near steady by 4 tsols.*

## 5. Comparison with Analytical Air-Sea Interaction Models

It is instructive to compare the range of results from mtWRF with the analytical model of M07. The mtWRF solutions provide an opportunity to explore both the validity of key assumptions made by M07, as well as the overall analytical solution space. In M07, an atmosphere with unchanging temperature is assumed to blow over a lake that can evaporate and cool via both a latent and sensible heat flux exchange with the atmosphere. A bulk eddy exchange formulation is used (Eq. 1), but with a fixed transfer coefficient that implicitly has no dependence on ambient conditions. Likewise, there is an implicit assumption that no circulation is present other than a mean wind blowing across the lake, and the wind speed is fixed in time. Finally, M07 seeks a steady solution that requires a balance between the sensible and latent heat fluxes (e.g., Bowen ratio = -1.0). In the M07 model, the sensible heat flux will necessarily increase monotonically in time, because the lake is cooling and the atmosphere temperature is static; the air temperature remains unaltered despite the nonzero sensible heat flux. The latent heat flux decreases monotonically in time due to the decrease in saturation vapor pressure associated with the falling lake temperature and the moistening atmosphere. Under the assumptions above, a steady-state solution where the sensible and latent heat fluxes are equal in magnitude but opposite in sign may be found from which an equilibrium lake temperature and lake evaporation rate may be determined. Under some conditions, an equilibrium solution is not physical, because the equilibrium lake temperature falls below the freezing point; the lake will freeze before flux equilibrium is reached.

The M07 assumptions are consistent with the mtWRF solutions only in some cases. In most simulations with a cooling lake, the temperature of the air blowing over the lake decreases quickly as soon as it encounters the lake. This is consistent with what is generally seen on Earth: Most of the cooling of air blowing over a cold lake occurs within the first several km of the coastline [Phillips, 1972]. Consistent with M07, it is true that the simulated lake cools more substantially than the air, especially for shallow mixed layers. Thus, to first order, the assumption by M07 of a static air temperature might be considered reasonable in some scenarios, and the sensible heat flux is often driven more by the large drop in the lake temperature than the comparatively smaller change in air temperature. The greatest violation of the constant air temperature scenario appears to be in cases where there is a background mean wind.

Buoyancy and sea breeze circulations develop in mtWRF, which alters the initial wind or creates a wind if there was none initially present. Thus, the wind is generally not constant in time and it also has substantial spatial structure and variation. The bulk exchange coefficient also changes with time due to changes in the bulk Richardson Number (i.e., static stability changes and wind shear changes). The magnitude of the sensible and latent flux often, but not always, comes close to equality, as in M07, especially with shallow mixed layers. In deeper mixed layer cases, the Bowen ratio tendency often flattens well before reaching -1.0, and the fluxes would only approach equality over very, very long time scales. Those timescales, however, are so long that they become comparable to or greater than seasonal timescales, and in that case the overall background conditions would change (e.g., seasonal changes in the thermal, moisture and wind state). Also like M07, many scenarios result in a lake temperature that is locally far below freezing regardless of whether an equilibrium flux condition is achieved.

It could be argued that although the flux balance condition of M07 is frequently achieved in mtWRF, there are scenarios where that balance solution is a trivial solution to the problem. The trivial case is where the turbulent fluxes become so small that they provide essentially no forcing (i.e., effectively zero fluxes). Even in the cases where a flux balance condition is not achieved, many of the solutions approximate the trivial solution. This trivial solution exhibits a very stable and moist marine boundary. The stable marine layer has weak winds, small bulk turbulent transfer coefficients, and reduced air-sea gradients in temperature and moisture. While latent heat flux may dominate, as indicated by the Bowen ratio, the actual fluxes are inconsequentially small. In this sense, the equilibrium flux solution of M07 becomes effectively valid with the caveat that fluxes and winds are near zero.

On Earth, an equilibrium between the turbulent fluxes is rarely observed. Bowen ratio measurements over lakes, oceans and wetlands are typically -0.5 to 0 [e.g., Roulet et al., 2986; Den Hartog et al., 1994; Vallet-Coulomb, 2001; Lenters et al., 2005; Elsawwaf et al., 2010] and are never observed to approach -1.0. Only over semi-arid land does the sensible heat flux equal the latent heat flux in magnitude [e.g., Goutorbe et al, 1994]. This reality holds from the tropics to the arctic with exceptions for extreme events where arctic air passes over warm lakes [e.g., Philips, 1972], which is not likely for Titan where the pole to equator temperature gradient is only a few Kelvins and local values are less [Tokano et al., 2005]. Thus, the mtWRF simulations with unequal fluxes are in concert with the overwhelming weight of terrestrial observations (to the extent that air-sea exchange on Earth serves as an analog to Titan): equilibrium is not required.

Both the buoyancy circulation and the sea breeze circulation can transport moisture vertically and horizontally. The buoyancy circulation, in particular, has the ability to export heat and moisture aloft, well beyond the horizontal scale of the lake itself. When viewed as a thermodynamic system, energy balance potentially requires consideration of a large boundary for the system rather than just between the bounding shoreline. Energy balance need not be achieved over the lake, let alone in a single atmospheric column over the lake. Latent and sensible heat fluxes over the lake can be balanced by exchanges far away. On Earth, when the net energy flux exceeds that available from radiative forcing and subsurface conduction, it is known as the Oasis effect [e.g., Holmes and Robertson, 1958; Linacre et al., 1970], and is usually associated with dry desert air encountering a lake or irrigated land. The Oasis effect may be amplified in these simulations, as the ground is considered to be completely dry. In most of the mtWRF cases, the secondary circulations that develop play a large role in the air-sea exchange, and the impact of the buoyancy and sea breeze circulations must be accounted for in addition to whatever exchanges may be driven by the initial mean wind. The atmosphere immediately above the lake does not continually moisten due to ongoing latent heat flux, although moisture is continually added to the atmosphere system. Instead, the moisture tendency from evaporation is offset by export of that moisture by the circulation.

Despite the substantial differences between the assumptions of M07 and the mtWRF assumptions, it is fair to say that M07 did accurately describe the qualitative nature of the problem. The lake does cool from evaporation (even if it is excruciatingly slow in some cases) and the atmosphere does moisten (even if that moistening is small). Equality of the fluxes can be achieved over a range of timescales, but owing to the change in air temperature, wind, static stability, atmospheric moistening and the associated variations in turbulent exchange, the state of the lake and atmosphere would be different from what M07 predicted. Accordingly, the rate of evaporation would be different, as would the threshold conditions for when a lake would freeze. The mtWRF solutions indicate that the details of the solution are highly dependent on a large number of parameters and initial conditions. It is very difficult (perhaps not even possible) to reduce the problem to a simple analytical model.

Depending on the initial condition, latent heat fluxes from mtWRF can be effectively zero or up to several hundred W/m$^2$. Without further constraints, the mtWRF results provide only upper and lower limits on lake evaporation rates equivalent to this range of latent heat flux. M07 indicate that a flux of 6 x 10$^{-4}$ kg m$^{-2}$ yr$^{-1}$ over the whole planet is required to balance the photochemical loss of methane. For a flux given in W/m$^2$ and the latent heat of vaporization, L, in Table 1, evaporation rates in units of kg m$^{-2}$ yr$^{-1}$ are given by:

$$\text{Evaporation (kg m}^{-2}\text{ yr}^{-1}\text{)} = 0.17 * \text{Flux (W m}^{-2}\text{)}. \qquad \text{Eq (3)}$$

Thus, as noted by M07, even a few W/m$^2$ and a lake coverage of a few percent is sufficient to balance the photochemical loss rate. For example, a flux of 1 W/m$^2$ with a global lake coverage of 1% is equivalent to a net global evaporation rate of 1.7 x 10$^{-3}$ kg m$^{-2}$ yr$^{-1}$, which is an order of magnitude larger than what is required.

The above calculations do not take into account precipitation. For the lower end of the estimated latent heat fluxes from mtWRF, and assuming that much of the vapor is returned somewhere to the surface, the net evaporation (evaporation minus precipitation) could, in fact, be close to the photochemical loss rate. The low end of the evaporation rates in mtWRF would not produce a noticeable change in lake levels. For the larger evaporation rates of several hundred W/m$^2$, it would be remarkable if the evaporation and precipitation could balance globally to within the ~10$^{-4}$ kg m$^{-2}$ yr$^{-1}$ needed to match the photochemical loss rate. In the absence of an exact balance, that would mean the atmospheric abundance of methane could fluctuate from Titan year to Titan year with amplitudes well above the average loss. Since the reservoir of methane in the atmosphere is large (~6000 kg/m$^2$) compared to the known surface reservoir, however, the atmospheric variation would be fractionally small. From the standpoint of the lakes, pumping as much as 50 kg m$^{-2}$ yr$^{-1}$ of methane over their surface would result in noticeable changes in lake level in the absence of resupply, as noted by M07.

## 6. Freezing Lakes and Swamps

Since many of the simulations show appreciable cooling of the lake, regardless of whether flux equilibrium is achieved, the possibility of lakes freezing is a reasonable consideration. The exact temperature at which freezing occurs would depend on the composition of the lakes. Stofan et al. [2007] indicate that the presence of dissolved nitrogen can depress the freezing temperature by 3 to 5 K below the nominal equatorial temperature of 93.6 K. Many simulations show it is possible to cool by more than 5 K in a reasonable time (i.e., fractions of a Titan season). Hoftgartner and Lunine [2013] find similar results depending on the fraction of ethane. Tokano [2009b] suggested that pure and shallow methane lakes could freeze seasonally, while the deeper and less pure lakes were more likely to remain liquid. All of these theoretical conclusions, however, make assumptions, at least implicitly, about the poorly understood density behavior of liquid methane and methane solutions near the triple point.

Based on the mtWRF simulations, the most conducive conditions for freezing are shallow mixed layers, low atmospheric humidity, and a non-zero background wind. Deep mixed layers are not conducive to freezing, and it is reasonable to conclude that lakes with deep mixed layers will not freeze, at least under the range of conditions explored herein. That is not to say that the freezing of deep lakes is completely excluded on Titan; they might, as long as the cooling fluid remains near the surface of the lake (i.e., the deep lakes have a shallow mixed layer). Deep lakes are also not deep everywhere. Radar bathymetry [Hayes et al., 2010; Wall et al., 2010; Mastrogiuseppe et al., 2014] indicates the lake bottoms have a relatively gentle slope from the shoreline. The depth of the mixed layer can be no deeper than lake itself at any given point, thus the areas near the shoreline may behave as shallow mixed layers that could be more susceptible to freezing, particularly if horizontal transport (i.e., currents) is minimal. If cold fluids are denser than warm fluids, the cooling of the shallows would necessarily produce a buoyancy-driven current with a surface return flow of warmer fluid directed toward the shoreline. Tokano and Lorenz (2016) indicated that the lake temperature is generally higher near the shoreline, which is likely explained by a combination of both currents and radiation, neither of which processes are currently represented in mtWRF.

Radar backscatter indicates very smooth lake surfaces right up to the shoreline (excluding "magic islands" [Hoftgarner et al., 2016]), which were interpreted to be liquid when dialectric properties and brightness temperature were considered [Stofan et al., 2007; Wye et al., 2009]. These results would seem to rule out ice covered lakes as a widespread phenomenon. Yet, given the exotic and poorly constrained properties of the lakes, ice cannot be completely ruled out for all lakes at all times; Cassini conducted flybys and not continuous global mapping. The smoothness of the lakes, if liquid, requires low wind speeds. Given Titan's sluggish circulation, low wind conditions are probable, but the mtWRF simulations show that evaporation from a liquid surface does generate local circulations that can be intense by Titan standards. If the largest lakes are indeed liquid then this places constraints on wind speed, which at least reasonably exclude initial conditions that generate the strongest simulated circulations. While the freezing of the large seas and lakes of Titan cannot be completely ruled out, the overwhelming consensus based on solid evidence is that the reservoirs are liquid. What is possibly more likely is the freezing of very small and shallow lakes or ponds, or infrequent events that might temporarily freeze at least part (the shallows) of a large lake.

It remains a curiosity that so few clouds were observed in the northern summer if the lakes are liquid [e.g., Turtle et al., 2018]. While the appearance of clouds nearby lakes is not guaranteed (see Section 8), liquid lakes should be an active source of methane vapor even with relatively low latent heat fluxes. General circulation models predicted deep convective clouds associated with the warm summer boundary layer and upwelling at the edge of a mean meridional polar cell or within the intertropical convergence zone [Mitchell et al. 2006; Newman et al., 2016], but these clouds failed to show themselves during any of the Cassini flybys [Schneider et al., 2012; Lora et al., 2015]. The mtWRF simulations show that the nearby lake environment can be stabilized due to the cold marine layer, but that effect appears to be localized. Perhaps the northern lake district, taken as a whole, acts to stabilize all of the northern high latitudes against deep convection and clouds in general. Barth and Rafkin [2007] found that moistening a Huygens-like thermal profile results in convective available potential energy (CAPE) that feeds deep convection. If the atmosphere is moistened while also cooling the lower levels, it is possible to remain convectively stable with no CAPE. The methane that is released in this scenario would then be transported globally, including to the southern high latitudes, where deep convection becomes possible half a Titan year later during southern summer. This is more likely if the southern high latitude surface is relatively dry so that heating results in low level heating, unlike a possibly damp northern high latitude surface.

Damp ground can also cool by evaporation, and unlike a lake it cannot mechanically mix. Instead, the thermal structure is controlled by the turbulent fluxes and conduction/diffusion processes with the subsurface. Although we do not model damp ground in this study, extrapolation of the lake simulations may provide guidance on the behavior. Swamp-like surfaces might behave somewhat like a lake with subsurface energy conduction replacing the mixing process. It is worth considering whether a damp surface could cool to the frost temperature or even produce a frozen crust. Early analysis of Huygens penetrometer data suggested that it landed on a crust that released methane and then failed mechanically after heating [Abbott, 2005]. This could be consistent with the melting of an ice crust and the release of methane vapor from the damp ground beneath; however, this preliminary interpretation has generally given way to an explanation of the signal that favors rocks and pebbles in conjunction with the kinetics of landing on a heterogeneous, bumpy surface. A moist or damp surface without a crust is still reasonable at the Hugygens landing site [Zarnecki et al., 2005; Atkinson et al., 2010]. If damp ground could freeze due to evaporation, the process may be especially favored in the cold polar night. If this happens near the marine layer environment, hoar frosts and ice-coated ground could be a regular feature in the polar marine microclimate.

## 7. Winds and Waves

Building on various previous predictions about the existence and dynamics of waves on Titan (e.g., Ghafoor et al. [2000] and Lorenz and Hayes [2012]), Hayes et al. [2013] analytically explored the conditions required to generate and grow capillary-gravity waves in Titan's lakes and seas. They determined the threshold wind speed required for wave generation as a function of lake composition. They quantified these threshold wind speeds using the wind as measured at 10 meters above the surface, which is denoted as $U_{10}$. The $U_{10}$ diagnostic is used to compare the threshold surface wind stress calculated by Hayes et al. [2013] and the winds predicted by mesoscale models and general circulation models. For a pure methane lake, Hayes et al. [2013] derived a threshold of $U_{10}^{th}$ = 0.4 m/s for the onset of wind-driven waves. By inverting equation 2.3 in Hayes et al. [2013], we derive the threshold surface friction velocity, $u_*^{th}$, to be 0.012 m/s. This surface friction velocity represents the speed of the atmosphere in the constant flux layer above the lake/sea surface. (We encourage future studies to employ u∗ or surface stress rather than $U_{10}$, since the prior are standard outputs for most models and are also standard parameters in micrometeorology).

Our canonical experiment (Simulation 4), discussed previously, provides a good example of how our simulated surface friction velocities compare to the $u_*^{th}$. Fig. 22 shows the value of the friction velocity for every horizontal grid point over the lake as a function of time for the canonical simulation. Overall, the winds rarely exceed the threshold wind velocity, with just intermittent bursts of wind that are strong enough to potentially generate waves. The bulk of these faster winds that exceed the threshold for wave generation occur over the first half of the first tsol, due to the initial buoyant methane plume that forms during the early phases of these idealized simulations. However, the mesoscale dynamics quickly evolve from a circulation dominated by a plume land breeze to a circulation dominated by a sea breeze, and thus the magnitude of the largest winds weakens over the course of the next 4 tsols.

Our simulations predict that wind gusts that exceed the threshold friction velocity are infrequent. This can be seen in Fig. 23, which compares the histogram of $u_*$ for tsol 1 versus tsol 5 of the simulation. In the tsol 1 histogram, a large number of the wind gusts exceed the threshold for wave generation. The tsol 1 winds, however, are part of the spin-up (plume phase) of the simulation and likely do not represent physically realistic or likely circulations.

The histogram of $u_*$ for tsol 5 of the simulation is likely more representative. During this stage of the simulation, a sea breeze circulation exists over the lake, with a small fraction of gusts exceeding the threshold for generating wind-driven waves. Fewer than 1000 out of 20,000 wind counts exceed the wave threshold (i.e., ~5%) at tsol 5, and these winds barely exceed the threshold speed (0.015 m/s vs the 0.012 m/s threshold). A reasonable interpretation is that this frequency should be sufficient to produce a wave somewhere, although it may not be sufficient to be observed if it is a very brief disturbance and/or highly confined in area and/or a very weak wave (due to the low wind speeds).  It's also worth noting that a slight shift of the threshold to higher winds (e.g. from 0.01 to 0.2 m/s) can drastically alter the statistics.  The uncertainty in the actual threshold for lakes must be considered as part of the interpretation of the modeled friction velocity distribution, as should the possibility that subgrid scale wind gusts could produce winds higher than those modeled here.

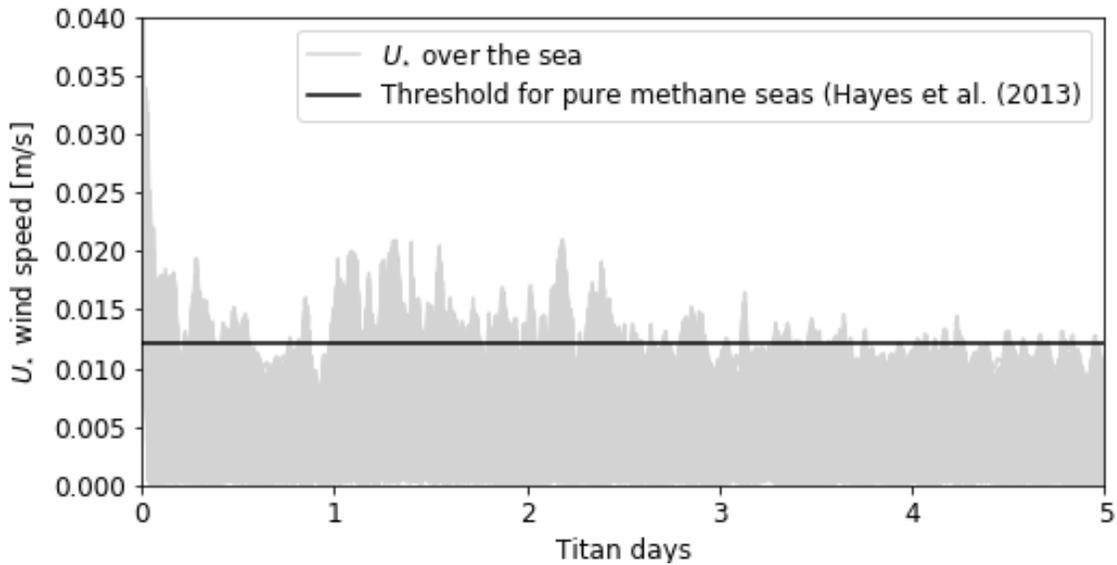

*Figure 22. All of the simulated wind speeds over the lake as a function of time for the canonical simulation (Sim 4). A grey line is plotted for each horizontal grid in the lake, 150 lines in total, which results in this collection of overlapping wind evolutions. The black line marks the threshold friction velocity for the generation of waves, based on Hayes et al. [2013].*

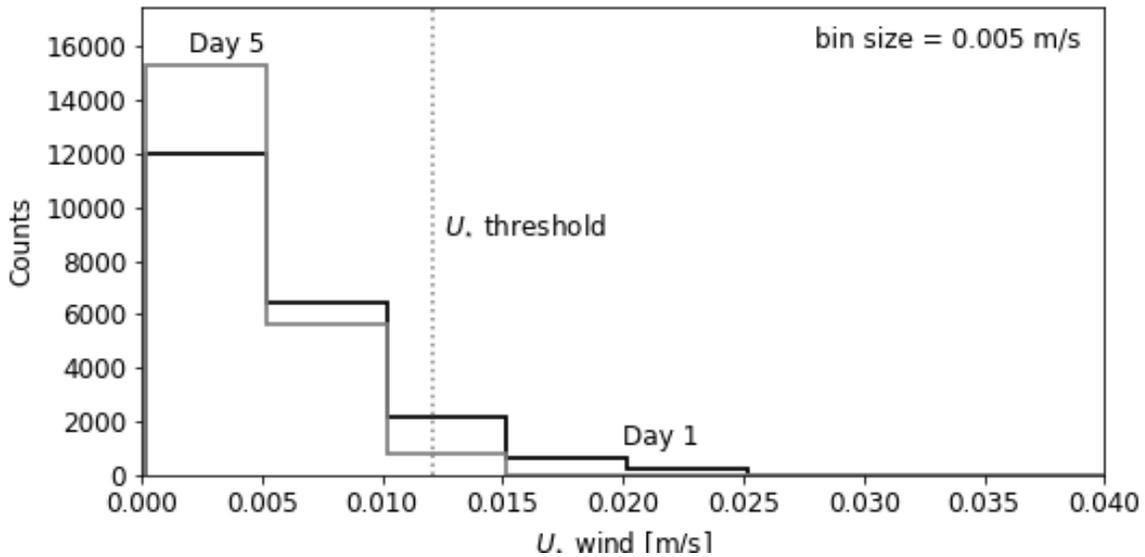

*Figure 23. Histograms at two different times in the canonical simulation show how the likelihood of exceeding the wind-drive wave threshold changes with time. The dashed vertical lines indicate the $U_*^{th}$ = 0.012 m/s threshold for the generation of wind-driven waves. The black line is the histogram of wind speeds for tsol 1 of the simulation and the grey line is the histogram of wind speeds for tsol 5 of the simulation.*

The existence of waves on Titan's seas and lakes remains unconfirmed. Cassini did infrequently observe surface features of the lakes of Titan that are consistent with wind driven waves [Barnes et al., 2014; Hofgartner. et al., 2014]. These surface features, however, are also consistent with other possible formation mechanisms, including floating and/or suspended solids and bubbles [Hofgartner. et al., 2014]. Therefore, there is insufficient evidence to definitively claim that the observed surface features are wind-driven waves.

Many simulations show winds that are stronger than those of Simulation 4. All of these are situations where a plume circulation dominates. In those cases, it becomes more likely that waves would appear, while it simultaneously remains true that there is no definitive observational evidence for widespread wave activity. This could mean that the initial conditions and configurations that result in strong and persistent plume circulations are not realistic, and that simulations with at least a sea breeze circulation that damps the plume circulation are more realistic. For those simulations with a sea breeze case, mtWRF suggests that waves driven by the mean wind are occasionally possible, especially given the uncertainty in wind thresholds. Wind gusts might be able to occasionally produce waves, but turbulent gusts decrease in magnitude as the marine layer strengthens in magnitude. The paucity of observational evidence that could be interpreted as wave activity strongly suggests that the conditions over the lakes of Titan are more likely to resemble the subset of mtWRF outcomes with strong and stable marine layers and minimal air-sea interaction. Absence of evidence is not evidence of absence, and the paucity of wave observations cannot definitively constrain the magnitude of air-sea circulations that occur on Titan.

# 8. Clouds

Microphysics are not active in the simulations, but we can evaluate if and when the model atmosphere approaches or exceeds saturation. None of the simulations initialized with a dry atmosphere come anywhere close to producing a saturated atmosphere. Even the most extreme evaporation cases are unable to generate an environment conducive to cloud formation. Generally speaking, the circulations export evaporated methane from the lake region, and the stronger the evaporation the more rapid the transport of vapor.

Before proceeding to discuss the most favorable cloud formation cases, it is essential to reestablish the historical and genetic cloud nomenclature used by the terrestrial community, but which has become muddled in the Titan literature. First, fog is a cloud that is in contact with the ground [American Meteorological Society, 2018]. "Ground fog" would then seem a redundant terminology, but instead refers to a patchy fog with mostly unclouded skies above [American Meteorological Society, 2018]. There are other types of fog that can be further identified, based primarily on the formation mechanism (e.g, mixing fog, radiation fog, frontal fog). While fog is a low cloud, low clouds are not necessarily fogs and the two terms should not be used interchangeably. The distinction is important, because the radiative and microphysical forcings in fog are distinctly different from other low clouds [Cotton et al., 2010]. Unfortunately, from a satellite perspective, fog and stratus (a type of low cloud) are often indistinguishable.

Second, "lake effect" clouds are not just any clouds caused by a lake. Lake effect clouds almost always refer specifically to the clouds generated when cold air blows over a warm lake accompanied by concomitant large turbulent energy fluxes.  The lake effect snowstorms in the Great Lakes region are the prime example of lake effect clouds [Holroyd III, 1971; Lavoie, 1972; Hjemfelt, 1990; Niziol et al., 1995]. Large sensible and latent heat fluxes produce shallow but intense convective clouds that can result in copious amounts of precipitation near the shoreline, particularly if the air is forced over rising terrain.  The reference to lake effect clouds in the Titan literature (e.g., Brown et al. 2009a), while colloquially accurate, seems unlikely to be referring to the meaning of lake effect clouds generally accepted by the terrestrial meteorological community.

There are sea breeze or lake breeze clouds, which should not be confused with lake effect clouds.  These are clouds generated by the upward forcing of moist air, possibly moistened by the lake or sea, along the sea or lake breeze convergence boundaries.  If the air is conditionally unstable, convective clouds may develop along the sea breeze, but not necessarily so.  A prime example of these types of clouds is found along the Florida peninsula [Pielke, 1974; Blanchard and López, 1985; Kingsmill, 1995].  It is likely that these are the types of clouds that have been improperly identified as lake effect clouds in the Titan literature.  We encourage future papers to more strictly follow terrestrial nomenclature and genetic descriptions of clouds to avoid further confusion.

Scenarios that start with a humid atmosphere are one place to look for the possibility of cloud formation. We explicitly exclude in our analysis the upper domain of the model where the atmosphere may be initialized to saturation (Fig. 4); those regions are conducive to cloud formation, but that situation is not due to air-sea interaction and is irrelevant to cloud formation mechanisms associated with the lake. In the stable 80% humidity cases, saturation conditions are found at initialization at altitudes above ~2 km. Below 2 km, saturation conditions are never achieved.  Thus, even in the most humid initial condition, the lake does not directly force cloud formation. The atmospheric relative humidity in the marine layer is usually *lower* than the initial 80% relative humidity, typically ranging from 50% to 75%. The circulations vertically transport the initial very moist atmosphere away from the surface, and the combination of vapor from evaporation and the developing net atmospheric circulation is unable to return the atmosphere to the initial state. In terms of total methane vapor, the atmosphere is continually gaining methane mass, but that vapor is unable to condense anywhere near the lake due to the net horizontal divergence of that vapor away from the lake .

In the unstable 80% case (sim #36), the atmosphere is initialized everywhere subsaturated, but a saturated layer forms at ~4 km within the first two hours of the simulation. This cloud layer, however, covers the whole domain and is not associated with any forcing from the lake. Instead, the cloud formation is due to the thermally unstable initial conditions that supports convection and subsequently transports moisture upward. The clouds would be best described as stratocumulus.

The observational lack of extensive cloudiness extending over lakes argues for relative humidity well below 80%. Further, if the background relative humidity were at 80%, deep convection would be expected for a Huygens-like temperature profile [Barth and Rafkin, 2007]. The only way to prevent such convection would be to stabilize the lower atmosphere through cooling. This cooling can be achieved by the air-sea interaction, but the model simulations indicate that this cooling also comes with a net drying with respect to relative humidity. Of course, if the lakes are frozen, that could also explain the general lack of observed clouds, but that situation has mostly been ruled out, as previously discussed.

The inability of the RH=80% simulations to produce saturation in the marine layer after the initial plume phase also argues that marine layer clouds should be rare. Thus, the model simulations are consistent with observations of almost universally cloud-free lakes. Lakes provide a source of vapor, but the ensuing dynamics associated with the air-sea interaction keep the atmosphere from saturating and producing clouds. The single instance of a cloud found near a lake (Brown et al., 2009a) could be an exception to this rule, or the appearance of the cloud by the lake could be coincidental and completely unrelated to air-sea interaction whatsoever.

The unstable lower relative humidity cases (sims #34 and #35) remain well below saturation. The 20% case has RH ~40% just above the sea breeze front, and the 50% case peaks at RH ~60%. The relative humidity in the marine layer above the lake is somewhat lower than the initial relative humidity. The lower value is due to the mixing of vapor through the boundary layer, which brings drier air toward the surface. The stable lower relative humidity cases (Sim #61 and #62) show very little change from the initial conditions, but unlike the unstable cases, the humidity above the lake increases very slightly.

A cold, deep mixed layer experiment with RH=80% (sim #52) was conducted to see if a more rapidly developing sea breeze circulation might allow the near-surface atmosphere to cool and moisten more than the baseline simulation. It did not. The near-surface relative humidity dropped to under 70%. An imposed background wind (1 m/s), shallow mixed layer case with RH=80% was then performed (sim #53) to investigate whether that combination of parameters might enhance the evaporation and cool the near-surface air to saturation. It did not. No configurations were found that could produce a saturated marine layer or clouds that were in any way directly attributable to the lake.

Based on the non-zero relative humidity simulations, we conclude that the neglect of condensation processes in the mtWRF simulations is reasonable. Except for the upper model domain where relative humidity is initialized to 100%, no clouds would be expected to form. Activating a microphysical scheme would do nothing except increase the integration time.

## 9. Importance of Model Numerics

Titan is a weakly forced system. The magnitude of the tendencies in the model prognostic variables when compared to machine precision and numerical error (e.g., errors in advection operators) can be orders of magnitude smaller for Titan than for planets like Earth or Mars. For example, the diurnal range in temperature for Earth can be 10 K or more, while Mars can approach 100 K. On Titan, 1 K or less may be typical. Yet, the machine precision and the accuracy of numerical operators do not change. Consequently, the magnitude of the errors for Titan can be a significant fraction of the magnitude of the forcing and prognostic variables.

Errors of 0.01 K may not matter much for Mars, but on Titan they can. Indeed, we initially found many cases of an initially cold lake for which the lake cooling tendency was too small to produce a change in the lake temperature (this was brought to our attention by the anonymous and keen reviewer). By going directly into the code, we were able to determine that the lake temperature tendency was non-zero, but small. When multiplied by the time step and then added to the lake temperature from the previous time step, floating point arithmetic resulted in no net change in the lake temperature. This numerical error is an energy leak in the system—latent heat is being removed from the lake and deposited into the atmosphere, but the lake energy is not changing. Our solution was to accumulate the lake temperature tendencies over a long enough time scale so that the value eventually becomes large enough to be represented by the floating precision (i.e., a split-time scheme). Going to double precision instead of the standard four-byte precision can help to reduce the issue, but it does not eliminate the general concern. Once the split-scheme for tendency accumulations was added, a difference in nearly all solutions was found when compared to the standard, non-split scheme. While the general solutions of sea breezes and plume circulations remained robust, the rate of lake cooling and the details of the circulations were noticeably changed. Numerics matter.

The most pronounced and initially obvious numerical issues were associated with the dynamical fields and not with the numerical precision and lake thermodynamics. In the purely symmetric scenarios (those without mean wind), the left side of the domain should be a mirror image of the right. In no simulation is this condition precisely true, and in a few cases is it strongly violated. Some good examples of asymmetry were noted in Fig. 2d, Fig. 15, and Fig. 19. Many of the asymmetries appear to be a direct result of how periodic boundary conditions are implemented. The rightmost boundary is arbitrarily assigned to the value on the leftmost side. This does satisfy the periodic boundary condition, but it is indeed arbitrary. The leftmost boundary could just as easily have been assigned the value of the rightmost boundary, in which case the asymmetry would be reversed. The appearance of any asymmetry is still due to numerical noise, as a perfect model would produce equal values at both boundaries, in which case an arbitrary mapping is inconsequential. But there are no perfect models, and the arbitrariness is not inconsequential. For example, if the outflow of a plume circulation reaches the left boundary a little before the right boundary, the solution in the left half of the domain will bleed more and more into the right domain and produce an asymmetry. Given enough time, the entire solution may become nonphysical. This puts an inherent time limit on the total integration time. We have examined all the simulations to ensure that the results presented here are at integration times that show reasonably symmetry (unless otherwise noted).

Numerical instabilities were also found for the 1 m/s background wind case with open boundary conditions (sim #65; Fig. 20d). In this case, the solution was expected to be asymmetric, but the appearance of domain-wide momentum directed completely opposite to the initial mean wind cannot be physically correct. Once again, the boundary condition appears to play a role, because the momentum reversal appeared shortly after the plume outflow reached the domain boundary and the instability amplified quickly. Interestingly, the stronger background wind case of 3 m/s did not exhibit this behavior. Apparently, the stronger wind was of sufficient magnitude so that boundary condition inaccuracies were unable to flip the sign of the wind field (i.e., small, non-physical perturbations in a 3 m/s wind still results in a positive wind). It is also interesting to note that Simulation 56 with a 1 m/s background wind and open boundary conditions did not exhibit instability. Apparently, the persistent strong plume forcing associated with the deep mixed layer was sufficient to overwhelm numerical instabilities.

The issues with numerics lead to the very serious question of whether these simulations and simulations of any weakly forced planetary system can be trusted at all. Embedded within this issue is a lesson about blindly taking perfectly adequate and proven Earth models and applying them to alien worlds that occupy an atmospheric parameter space far outside terrestrial norms. One might even call into question the results of Earth models in regions with weak forcing. This paper would likely have been completed more than two years ago had we started with a WRF code base rather than our existing TRAMS model [Barth and Rafkin, 2007]. We found numerical errors and conservation issues in TRAMS that were too large for the air-sea interaction problem and spent considerable time trying to understand and (unsuccessfully) correct the errors. Often the TRAMS errors were manifested in a very rapid growth of strong, unidirectional (asymmetric) winds alternating in height. Like mtWRF, these nonphysical modes could be further amplified by the numerical boundary conditions (e.g., zero gradient or constant gradient lateral boundaries). TRAMS remains a perfectly adequate model for deep convective cloud simulations, because the diabatic convective forcings are enormous—comparable to Earth—and dominate by orders of magnitude over the background numerical noise.

After more than a year of attempting to unsuccessfully reduce TRAMS numerical artifacts to an acceptable level, we decided to try WRF, which was a considerably more modern and, hopefully, quieter numerical core. This turned out to be the case, and after considerably more effort of testing and documenting the growth and behavior of asymmetries in the solution, we became increasingly confident in the model solutions. Nearly all the simulations, if run long enough, eventually become substantially asymmetric and non-physical. For all the simulations from which we've derived conclusions, the asymmetric component (i.e., nonphysical solution) is small. Nevertheless, there are some mtWRF configurations that have large inherent instability, as we've noted. Additionally, mtWRF had to be modified to incorporate split time scheme in order to properly accumulate the very small turbulent tendencies.

It may very well be that running TRAMS or mtWRF as traditional mesoscale models with imposed GCM boundary conditions will help solve the numerical problems. The imposition of physical boundary conditions should keep the model from running toward nonphysical solutions. We expect to implement the 3-D, GCM-forced simulations soon and we will test this hypothesis.

The ultimate solution is to work with model core developers early in the design stage to ensure that the model numerics better deal with weakly forced conditions, including the instances that might be encountered on Earth. This problem is not just a Titan air-sea problem. Similar problems can plague Venus and Titan GCMs where the forcing on the angular momentum budget can be very small—perhaps comparable to numerical error [e.g., Read and Lebonnois, 2018]. Anytime a process is regularly producing small tendencies compared to the prognostic variable, or where the final forcing is the residual of large terms, there is potential for problems.

## 10. Summary and Conclusions

The terrestrial WRF model was modified to study air-sea interaction on Titan under a variety of idealized scenarios in two dimensions. The effect of lake size, lake mixed-layer depth, lake-atmosphere temperature differences, atmospheric humidity, and mean background wind were all investigated. The general, quasi-steady solution is a non-linear superposition of a plume circulation driven by the buoyancy of evaporated methane and a sea (or lake) breeze circulation driven by the thermal contrast between the cold marine layer over the lake and the warmer inland air. The specific solution depends on the configuration and ranges from a persistent and strong plume circulation with little to no sea breeze tor a rapidly developing sea breeze and highly suppressed plume circulation.

Solutions favoring a plume circulation are those where the lake temperature cools slowly or not at all. Hence, deep mixed layers are most associated with plumes. Simulations dominated by plumes tend to have the strongest winds, but the general lack of wave activity on Titan (suggesting low winds) as inferred by radar strongly suggests that strong plume circulations are unlikely. The most rapid lake cooling is associated with initial plume circulations. In many cases, the lakes cool to temperatures that are close to or possibly below the uncertain freezing temperature. Frozen lakes have not been observed. All of the model results taken together with observational constraints strongly suggest that widespread, strong plume circulations are unlikely on Titan. This then implies that there must be some semblance of a sea breeze circulation to counter the plume circulation tendency, and this in turn implies the presence of a cold marine layer that is needed to drive that sea breeze circulation.

Conditions favoring a damped plume circulation or an outright sea breeze are cold lakes and shallow mixed layers. Initial background wind, relative humidity and lake size have a minor impact on this general finding. If the lake remains near the temperature of the air, a plume circulation may dominate. It was found that an air-sea differential of ~2 K marked the transition from a plume-dominant circulation to a sea breeze-dominant circulation. Hence, if plumes are not present on Titan, it can be inferred that the lakes are at least ~2 K colder than the air.

Sensible and latent heat fluxes can rapidly vary in magnitude while their ratio (the Bowen ratio) remains remarkably stable. The largest fluxes are generally associated with a plume circulation since the wind speeds are strongest in those cases. Once a marine layer develops, the increased static stability and lower wind speeds are less favorable for vigorous turbulent exchange. If we restrict the analysis to simulations without a strong plume circulation, without excessively cold lakes, and without winds speeds sufficient to trigger waves, the highest magnitude fluxes may be considered unrealistic under current observational constraints.

The result from Simulation 61 is an example of a solution consistent with observational constraints (50% RH, 1 m mixed layer). Although the lake starts at the same temperature as the air, the lake cools quickly to the approximate 2 K threshold needed for the suppression of the plume circulation. Wind speed drops dramatically, and sensible and latent heat fluxes are -2 W/m$^2$ and ~6 W/m$^2$, respectively. This would lead to global evaporation rates that are small, but still sufficient to balance photochemical loss rates. That same latent heat flux, however, would produce only a 0.2 cm drop in lake level per year assuming no other sources or sinks.

Unfortunately, Cassini measurements alone are unlikely to provide strong constraints on the magnitude of the fluxes. Lake levels changes, if observed, are not only due to evaporation, but precipitation and possible resupply from surface and subsurface flows. An observed decrease in lake level could be due to evaporation, or it could be due to some other process entirely. Further, even if evaporation is enormous, the lake levels could be maintained by recharge. Finally, if the latent heat fluxes are small, as in Simulation 61, then lake level changes are not good indicators unless measurement accuracy is better than ~1 cm/year. Additional analysis of radar brightness temperatures would be extremely helpful. Jennings et al. [2011] suggested that lake temperatures should be colder in the spring and summer due to differences in thermal mass between the land and the seas, and there was some indication of this in their analysis of radar-derived surface temperatures.

If the smaller magnitude fluxes are realistic, then radiative processes should be considered in future work. Insolation on Titan is on the order of a few W m$^{-2}$, which is comparable to the turbulent fluxes predicted in this study. For a lake surface radiating upwards at 89K and an atmosphere radiating down at 94K, the net IR flux is almost 1 W m$^{-2}$. Both short and longwave radiative fluxes could contribute in a significant way to the net energy balance, which would in turn impact the air-sea exchange. For example, a lake with a shallow mixed layer might behave more like a deep mixed layer if there is net radiative heating that partially offsets cooling from the latent heat flux. More detailed simulations with a radiative scheme for the atmosphere and surface is an area for future investigation.

Background wind shifts and distorts the plume and sea breeze circulations in ways that are expected based on terrestrial analogs. Sea breeze fronts that oppose that mean wind tend to be sharp and strong while those moving in the direction of the mean wind are more diffuse. A 3 m/s wind is sufficient to keep the impact of air-sea interactions restricted to the downwind direction. With a 1 m/s wind, the plume circulation and the sea breeze can be of sufficient strength to propagate their effects upwind. In the cases where a cold and stable marine layer develops, the underlying lake is effectively shielded from the influence of the background wind.

The lake dimensions did not strongly impact the overall solution. The dimensions and to a lesser degree the strength of the plume and sea breeze circulations scaled with the lake. There is certain to be some size limit below which the scaling fails. For example, puddles may produce an evaporative plume but are unlikely to force much of a sea breeze. We did not investigate the limiting size, but based on Earth it is likely to be O(1 km), since this is the fetch over which air typically adjusts to the underlying lake condition [e.g., Phillips, 1972].

Many, but not all the simulations were found to achieve the flux balance assumed in M07. In our results, the air does cool, and the development of an atmospheric circulation impacts the initial wind. Changes in atmospheric stability also play an important role in determining the fluxes. Unlike M07, this study provides little constraint on the actual methane source rate from lakes unless the low flux simulations are assumed to be the most representative of reality.

The 2 km resolution mtWRF simulations may be of use in parameterizing air-sea interaction in general circulation models (GCMs). Titan GCMs can use similar bulk transfer turbulent closures [e.g., Lora and Ádámkovics, 2017], but the resolution of the global models are insufficient to fully capture the dynamics of sea breezes or plume circulations. It is the non-linear sum of the large-scale circulation, which is roughly the GCM solution plus the lake-induced circulations, which the GCM cannot resolve that determine the net surface exchange. Further, it is not clear that the virtual buoyancy effect is considered in the GCM evaporation schemes, which is important.

None of the simulations were able to produce through air-sea interaction an environment conducive to cloud formation. This is consistent with the very infrequent cloud observations from Casinni [e.g., Turtle et al., 2018]. The observed cloud asserted to be associated with a lake [Brown et al., 2009a] is inconsistent with any of the model configurations that were investigated. The observed cloud is almost certainly not a lake effect cloud, as commonly referred to in the terrestrial literature, but a causal formation mechanism tied directly to air-sea interaction cannot be ruled out based on the simulations presented here. There may be some reasonable range of air and lake configurations that could favor cloud formation. Conversely, the appearance of a cloud near a lake could be entirely coincidental.

Air-sea interaction on Titan depends on a great number of factors, and the physics is complex and nonlinear.  As complex as the model physics in this study already are, radiative effects and three-dimensional circulations driven by irregular coastlines and topography have not been considered.  Future work is needed to explore how the shape of coastlines and surrounding topography can influence the atmospheric circulation.  The simulations have also shown that lake circulations, particularly the effective lake mixing depth, are extremely important.  Ideally, future work would involve coupling a dynamic lake model to the atmospheric model.  The uncertainty in lake composition and the physical behavior of liquid and ice remains problematic, but parameteric sweeps could be done with the lake model in a manner similar to what has been done here.

Numerical stability and model precision were found to be important, particularly in the most weakly forced cases.  A few simulations became numerically unstable and produced non-physical wind regimes, while a few retained some semblance of reality but had troubling asymmetries.  Numerical boundary conditions played a large role in the instabilities.  The boundary condition issues may be resolved by imposing realistic properties derived from a GCM.  Initial attempts of the simulations using TRAMS were unsuccessful due to numerical issues, and this motivated the use of mtWRF, which turned out to have lower magnitude nonphysical computational modes.  Application of well-tested and proven terrestrial models to extraterrestrial conditions can be problematic, and care needs to be taken to ensure solutions are physically consistent while nonphysical signals are acceptably small.  The weak forcing on Titan required the implementation of time-splitting schemes to properly account for turbulent flux tendencies.

As is always the case, more and higher fidelity observations would be beneficial.  A mission like the Titan Mare Explore (TiME) previously proposed to the NASA Discovery Program would be the most direct way to obtain the measurements and observational constraints needed to quantify air-sea interaction on Titan [Stofan et al., 2013]. The Dragonfly mission recently selected under the NASA New Frontiers exploration program is destined for the tropical dunes of Selk Crater, but the meteorological measurements would still be useful in providing constraints, especially if there is moist ground exchanging methane with the atmosphere. Direct measurement of wind, temperature, and moisture profiles coupled to surface temperature and surface properties can provide strong constraints on turbulent fluxes, and these can be extrapolated to conditions over liquid reservoirs.

## Acknowledgements

This work was primarily supported through the NASA Planetary Atmosphere Program with important contributions from Southwest Research Institute and the substantial investment of the authors' personal time. Dr. Claire Newman and one anonymous reviewer provided enormously helpful comments and were responsible for identifying additional and important inconsistencies in the numerics that the original manuscript failed to identify. For this, the authors are extremely grateful.

Alexander, M.A., cBladé, I., Newman, M., Lanzante, J.R., Lau, N.C. and Scott, J.D., 2002. The atmospheric bridge: The influence of ENSO teleconnections on air–sea interaction over the global oceans. *Journal of Climate*, *15*(16), pp.2205-2231.

American Meteorological Society, 2018. Glossary of Meteorology. [Available online at http://glossary.ametsoc.org/wiki.

Atkins, N.T. and Wakimoto, R.M., 1997. Influence of the synoptic-scale flow on sea breezes observed during CaPE. Monthly weather review, 125(9), pp.2112-2130.

Atkinson, K.R., Zarnecki, J.C., Towner, M.C., Ringrose, T.J., Hagermann, A., Ball, A.J., Leese, M.R., Kargl, G., Paton, M.D., Lorenz, R.D. and Green, S.F., 2010. Penetrometry of granular and moist planetary surface materials: Application to the Huygens landing site on Titan. *Icarus*, *210*(2), pp.843-851.

Atreya, S.K., Adams, E.Y., Niemann, H.B., Demick-Montelara, J.E., Owen, T.C., Fulchignoni, M., Ferri, F. and Wilson, E.H., 2006. Titan's methane cycle. *Planetary and Space Science*, *54*(12), pp.1177-1187.

Barnes, J.W., Soderblom, J.M., Brown, R.H., Soderblom, L.A., Stephan, K., Jaumann, R., Le Mouélic, S., Rodriguez, S., Sotin, C., Buratti, B.J. and Baines, K.H., 2011. Wave constraints for Titan's Jingpo Lacus and Kraken Mare from VIMS specular reflection lightcurves. *Icarus*, *211*(1), pp.722-731.

Barnes, J.W., Sotin, C., Soderblom, J.M., Brown, R.H., Hayes, A.G., Donelan, M., Rodriguez, S., Le Mouélic, S., Baines, K.H. and McCord, T.B., 2014. Cassini/VIMS observes rough surfaces on Titan's Punga Mare in specular reflection. *Planetary science*, *3*(1), p.3.

Barth, E.L. and Rafkin, S.C., 2007. TRAMS: A new dynamic cloud model for Titan's methane clouds. *Geophysical Research Letters*, *34*(3).

Bird, M.K., Allison, M., Asmar, S.W., Atkinson, D.H., Avruch, I.M., Dutta-Roy, R., Dzierma, Y., Edenhofer, P., Folkner, W.M., Gurvits, L.I. and Johnston, D.V., 2005. The vertical profile of winds on Titan. *Nature*, *438*(7069), p.800.
94